\documentclass[conference]{IEEEtran}
\IEEEoverridecommandlockouts

\usepackage[T1]{fontenc}

\usepackage{cite}
\usepackage{amsmath,amssymb,amsfonts}
\usepackage{algorithmic}
\usepackage{graphicx}
\usepackage{textcomp}
\usepackage{xcolor}

\usepackage{orcidlink}

\usepackage[svgnames,table]{xcolor}
\definecolor{darkblue}{rgb}{0, 0, 0.7}
	
\usepackage{ifthen}
\usepackage{amssymb}

\usepackage[normalem]{ulem} 

\newcommand{\ext}[1]{}
\usepackage{svg}  

\usepackage{verbatim}
\usepackage[capitalise,noabbrev]{cleveref}

\crefname{lstlisting}{listing}{listings}
\Crefname{lstlisting}{Listing}{Listings}
\Crefname{figure}{Fig.}{Figs.}

\usepackage{listings}

\DeclareFixedFont{\ttb}{T1}{txtt}{bx}{n}{12} 
\DeclareFixedFont{\ttm}{T1}{txtt}{m}{n}{12}  

\usepackage{color}
\definecolor{deepblue}{rgb}{0,0,0.5}
\definecolor{deepred}{rgb}{0.6,0,0}
\definecolor{deepgreen}{rgb}{0,0.5,0}

\definecolor{maroon}{cmyk}{0, 0.87, 0.68, 0.32}
\definecolor{halfgray}{gray}{0.55}
\definecolor{ipython_frame}{RGB}{207, 207, 207}
\definecolor{ipython_bg}{RGB}{247, 247, 247}
\definecolor{ipython_red}{RGB}{186, 33, 33}
\definecolor{ipython_green}{RGB}{0, 128, 0}
\definecolor{ipython_cyan}{RGB}{64, 128, 128}
\definecolor{ipython_purple}{RGB}{170, 34, 255}
\definecolor{dartwin_purple}{RGB}{127, 0, 85}

\lstdefinelanguage{dartwin}{
    belowcaptionskip=0.25cm,
    morekeywords= {
        abstract,
        allocate,
        allocation,
        attribute,
        connect,
        constraint,
        def,
        default,
        connect,
        connection,
        doc,
        import,
        library,
        metadata,
        meta,
        occurrence,
        package,
        part,
        port,
        private,
        requirement,
        to,
        variation,
        variant,
    },
    alsoletter={\#},
    morekeywords=[2]{
        \#dartwin,
        \#digitaltwin,
        \#twinsystem,
        \#goal,
        \#vs,
        \#dartrans,
        \#dartwin_core,
        \#dartwin_before,
        \#dartwin_after
    },%
    sensitive=true,%
    morecomment=[l]{//},%
    morecomment=[s]{/*}{*/},%
    morestring=[b]',%
    morestring=[b]",%
    morestring=[s]{'''}{'''},
    morestring=[s]{"""}{"""},
    morestring=[s]{r'}{'},
    morestring=[s]{r"}{"},%
    morestring=[s]{r'''}{'''},%
    morestring=[s]{r"""}{"""},%
    morestring=[s]{u'}{'},
    morestring=[s]{u"}{"},%
    morestring=[s]{u'''}{'''},%
    morestring=[s]{u"""}{"""},%
    literate=
    {á}{{\'a}}1 {é}{{\'e}}1 {í}{{\'i}}1 {ó}{{\'o}}1 {ú}{{\'u}}1
    {Á}{{\'A}}1 {É}{{\'E}}1 {Í}{{\'I}}1 {Ó}{{\'O}}1 {Ú}{{\'U}}1
    {à}{{\`a}}1 {è}{{\`e}}1 {ì}{{\`i}}1 {ò}{{\`o}}1 {ù}{{\`u}}1
    {À}{{\`A}}1 {È}{{\'E}}1 {Ì}{{\`I}}1 {Ò}{{\`O}}1 {Ù}{{\`U}}1
    {ä}{{\"a}}1 {ë}{{\"e}}1 {ï}{{\"i}}1 {ö}{{\"o}}1 {ü}{{\"u}}1
    {Ä}{{\"A}}1 {Ë}{{\"E}}1 {Ï}{{\"I}}1 {Ö}{{\"O}}1 {Ü}{{\"U}}1
    {â}{{\^a}}1 {ê}{{\^e}}1 {î}{{\^i}}1 {ô}{{\^o}}1 {û}{{\^u}}1
    {Â}{{\^A}}1 {Ê}{{\^E}}1 {Î}{{\^I}}1 {Ô}{{\^O}}1 {Û}{{\^U}}1
    {œ}{{\oe}}1 {Œ}{{\OE}}1 {æ}{{\ae}}1 {Æ}{{\AE}}1 {ß}{{\ss}}1
    {ç}{{\c c}}1 {Ç}{{\c C}}1 {ø}{{\o}}1 {å}{{\r a}}1 {Å}{{\r A}}1
    {€}{{\EUR}}1 {£}{{\pounds}}1
    {::}{{{\textbf{::}}}}1
    {^}{{{\color{ipython_purple}\^{}}}}1
    {=}{{{\color{ipython_purple}=}}}1
    {+}{{{\color{ipython_purple}+}}}1
    {*}{{{\color{ipython_purple}$^\ast$}}}1
    {/}{{{\color{ipython_purple}/}}}1
    {+=}{{{+=}}}1
    {-=}{{{-=}}}1
    {*=}{{{$^\ast$=}}}1
    {/=}{{{/=}}}1
    {check()}{{{\color{black}{check()}}}}1,
    literate=
     *{-}{{{\color{ipython_purple}-}}}1
     {--}{{{\color{ipython_purple}\textendash}}}2
     {?}{{{\color{ipython_purple}?}}}1
     {\&}{{{\color{ipython_purple}\&}}}1
     {|}{{{\color{ipython_purple}|}}}1,
    tabsize=2,
    identifierstyle=\color{black}\ttfamily,
    commentstyle=\color{ipython_green}\ttfamily,
    stringstyle=\color{ipython_red}\ttfamily,
    keepspaces=true,
    showspaces=false,
    showstringspaces=false,
    rulecolor=\color{ipython_frame},
    frame=single,
    frameround={t}{t}{t}{t},
    framexleftmargin=2em,
    xleftmargin=2em,
    postbreak=\mbox{\textcolor{blue}{$\hookrightarrow$}\space},
    numbers=left,
    numberstyle=\tiny\color{halfgray},
    basicstyle=\scriptsize,
    keywordstyle={\color{dartwin_purple}\ttfamily\bfseries},
    keywordstyle=[2]{\color{dartwin_purple}\ttfamily\bfseries},
}

\newcommand{\dartwinInline}[2][\small]{\lstinline[language=dartwin,basicstyle=#1,breaklines=false]{#2}}

\usepackage[
]{todonotes}

\usepackage{tikz}

\usetikzlibrary{arrows}
\usetikzlibrary{decorations.markings}
\usetikzlibrary{arrows.meta}  
\usetikzlibrary{positioning}  
\usetikzlibrary{calc}
\usetikzlibrary{shapes}  
\usetikzlibrary{fit}  
\usepgflibrary{plotmarks} %

\usepackage{tikz-layers}

\definecolor{HighlightColor}{HTML}{FFC107}

\makeatletter
\pgfdeclareshape{labelpentagon}{
  \inheritsavedanchors[from=rectangle] 
  \inheritanchorborder[from=rectangle]
  \inheritanchor[from=rectangle]{center}
  \inheritanchor[from=rectangle]{north}
  \inheritanchor[from=rectangle]{north east}
  \inheritanchor[from=rectangle]{east}
  \inheritanchor[from=rectangle]{south east}
  \inheritanchor[from=rectangle]{south}
  \inheritanchor[from=rectangle]{south west}
  \inheritanchor[from=rectangle]{west}
  \inheritanchor[from=rectangle]{north west}
  \backgroundpath{
    \southwest \pgf@xa=\pgf@x \pgf@ya=\pgf@y
    \northeast \pgf@xb=\pgf@x \pgf@yb=\pgf@y
    \pgf@xc=\pgf@xb \advance\pgf@xc by-5pt 
    \pgf@yc=\pgf@ya \advance\pgf@yc by+5pt
    \pgfpathmoveto{\pgfpoint{\pgf@xa}{\pgf@ya}}
    \pgfpathlineto{\pgfpoint{\pgf@xa}{\pgf@yb}}
    \pgfpathlineto{\pgfpoint{\pgf@xb}{\pgf@yb}}
    \pgfpathlineto{\pgfpoint{\pgf@xb}{\pgf@yc}}
    \pgfpathlineto{\pgfpoint{\pgf@xc}{\pgf@ya}}
    \pgfpathclose
 }
}
\makeatother

\tikzset{%
sysml-framelabel/.append style={labelpentagon,draw,fill=white,anchor=north west,
    align=left,font=\sffamily\fontsize{8pt}{8pt}\selectfont,
},
sysml-redefined/.append style={-{open triangle 60}, postaction={decorate,
   decoration={markings,mark=at position -.65em with {\arrow[thick]{|};}}}
},
sysml-defined/.append style={-{Triangle[open,fill=white,length=1em,width=1em]}, postaction={decorate,
   decoration={markings,mark=at position -1.3em with {
    \node[rotate=\pgfdecoratedangle]{\Large\textbf{:}};
   }}}
},
sysml-inherit/.append style={-{Triangle[open,fill=white,length=1em,width=1em]}},
sysml-partof/.append style={-{Diamond[length=1.5em]}},
sysml-whitepart/.append style={-{Diamond[open,fill=white,length=1.5em]}},
sysml-reference/.append style={-{Diamond[open,fill=white,length=1.5em]}},
sysml-dotted/.append style={-{Stealth[scale=1]},ultra thick,dotted}
}

\tikzset{
    every node/.append style={align=center,font=\sffamily},
}

\tikzset{
block/.append style={align=center,rectangle,draw=black},
dartwinElement/.append style={draw=black,fill=white,font=\sffamily\footnotesize},
biglink/.append style={draw=black,very thick,-{Triangle[open,fill=white,length=5mm, width=5mm]}},
goal/.append style={dartwinElement,
    trapezium, trapezium left angle=75, trapezium right angle=75, 
    inner ysep=.75em, minimum height=3em,
    font=\sffamily\scriptsize,
    },
digitaltwin/.append style={dartwinElement, 
    rectangle, rounded corners=.5em,
    inner sep=1em,minimum width=3cm},
arbiter/.append style={langElement,
    shape=regular polygon, regular polygon sides=8,
    inner sep=0em},
actualtwin/.append style={langElement, rectangle,
    inner sep=1em,minimum width=5em},
port/.append style={dartwinElement,rectangle,
    font=\sffamily,
    inner sep=0em,minimum width=1em,minimum height=1em,
},
portlabel/.append style={align=center,font=\sffamily\scriptsize},
dtsystem/.append style={block,fill=gray!5,inner sep=0pt},
goal-relation/.append style={{Triangle[length=3mm, width=3mm]}-{Triangle[length=3mm, width=3mm]},thick},
goal-conflict/.append style={{Triangle[length=3mm, width=3mm]}-{Triangle[length=3mm, width=3mm]},thick,double},
goal-specialization/.append style={-{Triangle[open,fill=white,length=3mm, width=3mm]},thick},
flow/.append style={-{Stealth[length=2mm, width=2mm]},thick},
relationToGoal/.append style={-{Stealth[length=3mm, width=3mm]},dotted,thick},
}

\makeatletter
\pgfdeclareshape{dartwinLabelPentagon}{
  \inheritsavedanchors[from=rectangle] 
  \inheritanchorborder[from=rectangle]
  \inheritanchor[from=rectangle]{center}
  \inheritanchor[from=rectangle]{north}
  \inheritanchor[from=rectangle]{north east}
  \inheritanchor[from=rectangle]{east}
  \inheritanchor[from=rectangle]{south east}
  \inheritanchor[from=rectangle]{south}
  \inheritanchor[from=rectangle]{south west}
  \inheritanchor[from=rectangle]{west}
  \inheritanchor[from=rectangle]{north west}
  \backgroundpath{
    \southwest \pgf@xa=\pgf@x \pgf@ya=\pgf@y
    \northeast \pgf@xb=\pgf@x \pgf@yb=\pgf@y
    \pgf@xc=\pgf@xb \advance\pgf@xc by-5pt 
    \pgf@yc=\pgf@ya \advance\pgf@yc by+5pt
    \pgfpathmoveto{\pgfpoint{\pgf@xa}{\pgf@ya}}
    \pgfpathlineto{\pgfpoint{\pgf@xa}{\pgf@yb}}
    \pgfpathlineto{\pgfpoint{\pgf@xb}{\pgf@yb}}
    \pgfpathlineto{\pgfpoint{\pgf@xb}{\pgf@yc}}
    \pgfpathlineto{\pgfpoint{\pgf@xc}{\pgf@ya}}
    \pgfpathclose
 }
}
\makeatother

\tikzset{
separationLine/.append style={draw,ultra thick},
pics/frameSeparation/.style args={#1/#2}{code={  
    \draw[separationLine] 
        ($(#1.west |- #2.south)+(1em,-1em)$) -- 
        ($(#1.east |- #2.south)+(-1em,-1em)$);
}},
framelabel/.append style={draw,dartwinLabelPentagon,fill=white,anchor=north west,
    align=left,font=\sffamily\fontsize{8pt}{8pt}\selectfont,
},
pics/darStarFrameFit/.style args={#1/#2/#3/#4}{code={
\begin{scope}[on background layer]
    \node[fit=#4,draw,inner sep=2em] (-box) {};
    \node[framelabel,inner xsep=.5em,inner ysep=.25em]  (-label) at (-box.north west) {\textbf{#1} #2 \ifstrempty{#3}{}{\textbf{based on} #3}};
\end{scope}
}},
pics/dartwinFrameFit/.style args={#1/#2/#3}{code={
    \pic {darStarFrameFit={dartwin/#1/#2/#3}};
}},
pics/dartransFrameFit/.style args={#1/#2/#3}{code={
    \pic {darStarFrameFit={dartrans/#1/#2/#3}};
}},
}

\tikzset{
highlight/.append style={draw=HighlightColor,very thick,fill=HighlightColor!20},
}

\tikzset{%
pics/comfort goal/.style={code={
    \node[goal,#1] (comfort goal) {\textbf{Warm Comfort}\\
        Room Temperature $t$\\
        $t$ in range [l,u]
    };    
}},
pics/freezeprotect goal/.style={code={
    \node[goal,#1] (freezeprotect goal) {\textbf{No Freezing}\\
    Room Temperature $t$\\
    $t > 8$
    };    
}},
pics/savemoney goal/.style={code={
    \node[goal,#1] (savemoney goal) {\textbf{Saving Money}\\
    Energy cost $c$ \\
    Lower $c$ by\\
    considering temporal\\
    energy prices
    };    
}},
pics/energy goals/.style={code={
    \node[goal,#1] (-lowerenergy) {
        \textbf{Lower Energy}\\
        Energy $e$ [kWh]\\
        lower $e$ than before};
    \node[goal,#1,below right=.5 and .5 of -lowerenergy] (-whenpresent) {
        \textbf{Energy when present}\\
        Energy $e$ [kWh]\\
        Presence of people $p$\\
        Same $e$ as before};
    \node[goal,#1,below left=.5 and .5 of -lowerenergy] (-whenabsent) {
        \textbf{Energy when absent}\\
        Energy $e$ [kWh]\\ 
        Presence of people $p$\\
        Lower $e$ than before};
    \draw[goal-specialization,#1] (-whenabsent.north)  |- ($(-whenabsent.north)!.25!(-lowerenergy.south)$) -| (-lowerenergy);
    \draw[goal-specialization,#1] (-whenpresent.north) |- ($(-whenpresent.north)!.25!(-lowerenergy.south)$) -| (-lowerenergy);
}},
}

\tikzset {
pics/energysave dt/.style={code={
    \node[digitaltwin] (-dt) {Energy Saving};
    \node[port,label={[portlabel,below]{ }}] (-comfort-temp-out) at (-dt.east) {$\rightarrow$};
    \node[port,label={[portlabel,below]{ }}] (-comfort-temp-in) at (-dt.west) {$\rightarrow$};
    \node[port,label={[portlabel,below]{ }}] (-presence) at (-dt.south) {$\uparrow$};   
}}, 
pics/comfort dt/.style={code={
    \node[digitaltwin] (-dt) {Thermostat Logic};
    \node[port,label={[portlabel,below]{ }}] (-heater) at (-dt.east) {$\rightarrow$};
    \node[port,label={[portlabel,below]{ }}] (-comfort-temp) at (-dt.west) {$\rightarrow$};
    \node[port,label={[portlabel,below]{ }}] (-temp) at (-dt.south) {$\uparrow$};   
}}, 
pics/freezeprotect dt/.style={code={
    \node[digitaltwin,#1] (-dt) {Freeze Protection};
    \node[port,#1,label={below:{}}] (-temp) at (-dt.west) {$\rightarrow$};
    \node[port,#1,label={below:{}}] (-heater) at (-dt.south) {$\downarrow$};   
}}, 
}

\tikzset{
pics/noswing goal/.style={code={
    \node[goal,#1] (noswing goal) {\textbf{No Swinging}\\\textbf{Motion}\\ Angle $\theta$ \\ $\theta_{end\,time} = 0$};    
}},
pics/constraints goal/.style={code={
    \node[goal,#1] (constraints goal) {\textbf{Respect System}\\ \textbf{Constraints}\\ Trace $t$ \\No violations over $t$};    
}},
pics/minimize goal/.style={code={
    \node[goal,#1] (minimize goal) {\textbf{Minimize Trajectory}\\\textbf{Duration}\\ Trace duration $d$ \\ minimize($d$)};    
}},
pics/nocollision goal/.style={code={
    \node[goal,#1] (nocollision goal) {\textbf{Respect Dynamic}\\\textbf{ Position Constraints}\\ Trace $t$\\ No violations over $t$};    
}},
pics/validation goal/.style={code={
    \node[goal,#1] (validation goal) {\textbf{Continuous Validation}\\ Validation metrics $vm$\\$vm$ conform to thresholds};    
}},
pics/container goal/.style={code={
    \node[goal,#1] (container goal) {\textbf{Respect Container}\\\textbf{Kinetic Constraints}\\ Trace $t$\\No violations over $t$};    
}},
pics/trajectory dt/.style={code={
    \node[digitaltwin,#1,minimum width=4cm] (-dt) {Trajectory};
    \node[port,#1,label={below:{}}] (-position) at ($(-dt.south)!.75!(-dt.south west)$) {$\uparrow$};
    \node[port,#1,label={below:{}}] (-swing) at (-dt.south) {$\uparrow$};
    \node[port,#1,label={below:{}}] (-controllers) at ($(-dt.south)!.75!(-dt.south east)$) {$\downarrow$};   
}},
pics/wrap trajectory ports/.style={code={
    \node[port,label={[portlabel]below:{Motor\\Position}}] (-position) at (#1.south -| trajectory-dt-position) {$\uparrow$};
    \node[port,label={[portlabel]below:{Swing\\Angle}}] (-swing) at (#1.south -| trajectory-dt-swing) {$\uparrow$};
    \node[port,label={[portlabel]below:{Motor\\Controllers}}] (-controllers) at (#1.south -| trajectory-dt-controllers) {$\downarrow$};  
    \draw[flow] (-position) -- (trajectory-dt-position);
    \draw[flow] (-swing) -- (trajectory-dt-swing);
    \draw[flow] (trajectory-dt-controllers) -- (-controllers);
}},
}

\usepackage{xspace} 

\newcommand{\cf}{cf.\xspace}
\newcommand{\eg}{e.g.\xspace}

\newcommand{\ie}{i.e.\xspace}

\usepackage[acronyms,shortcuts,section=section,style=super,nopostdot,nogroupskip,nomain,nonumberlist]{glossaries}
\glsdisablehyper

\newacronym{adl}     {ADL}   {Architecture Description Language}
\newacronym{ai}     {AI}    {Artificial Intelligence}
\newacronym{api}    {API}   {Application Programming Interface}
\newacronym{at}     {AT}    {Actual Twin}
\newacronym{as}     {AS}    {Actual System}
\newacronym{cs}     {CS}    {Control System}
\newacronym{cps}    {CPS}   {Cyber-Physical System}
\newacronym{dsl}    {DSL}   {Domain-Specific Language}
\newacronym{ds}     {DS}    {Digital Shadow}
\newacronym{dspl}   {DSPL}  {Dynamic Software Product Line}
\newacronym{dt}     {DT}    {Digital Twin}
\newacronym{dts}    {DTS}   {Digital Twin System}
\newacronym{iot}    {IoT}   {Internet of Things}
\newacronym{mde}    {MDE}   {Model-Driven Engineering}
\newacronym[longplural={Properties of Interest}]
    {poi}    {PoI}   {Properties of Interest}

\newcommand{\devops}{DevOps\xspace}

\newcommand{\DT}{\ac{dt}\xspace}
\newcommand{\DTs}{\acp{dt}\xspace}
\newcommand{\DTS}{\ac{dts}\xspace}
\newcommand{\DTSs}{\acp{dts}\xspace}

\newcommand{\AT}{\ac{at}\xspace}

\newcommand{\DSL}{\ac{dsl}\xspace}
\newcommand{\DSLs}{\acp{dsl}\xspace}

\newcommand{\DarTwin}{\textsf{DarTwin}\xspace}
\newcommand{\DarTwinDSL}{\textsf{DarTwin}~\textsf{\DSL}\xspace}
\newcommand{\SysMLone}{SysML\;\!v1\xspace}
\newcommand{\SysML}{SysML\;\!v2\xspace}

\usepackage{amsthm}

\newtheorem*{remark}{Remark}

\begin{document}

\title{DarTwin made precise by \SysML~-- An Experiment}



\author{
\IEEEauthorblockN{{\O}ystein Haugen~{\hypersetup{pdfborder={0 0 0}}\orcidlink{0000-0002-0567-769X}}}
\IEEEauthorblockA{\textit{{\O}stfold University College} \\
Halden, Norway \\
{\hypersetup{pdfborder={0 0 0}}\href{mailto:oystein.haugen@hiof.no}{oystein.haugen@hiof.no}}}
\and 
\IEEEauthorblockN{Stefan Klikovits~{\hypersetup{pdfborder={0 0 0}}\orcidlink{0000-0003-4212-7029}}}
\IEEEauthorblockA{
\textit{Johannes Kepler University} \\
Linz, Austria \\
{\hypersetup{pdfborder={0 0 0}}\href{mailto:stefan.klikovits@jku.at}{stefan.klikovits@jku.at}}}
\and 
\IEEEauthorblockN{Martin Arthur Andersen~{\hypersetup{pdfborder={0 0 0}}\orcidlink{0009-0004-9991-3578}}}
\IEEEauthorblockA{\textit{{\O}stfold University College} \\
Halden, Norway \\
{\hypersetup{pdfborder={0 0 0}}\href{mailto:martin.a.andersen@hiof.no}{martin.a.andersen@hiof.no}}}
\and 
\IEEEauthorblockN{Jonathan Beaulieu~{\hypersetup{pdfborder={0 0 0}}\orcidlink{0009-0006-7383-8746}}}
\IEEEauthorblockA{\textit{École de technologie supérieure (ETS)} \\
Montréal, Canada \\
{\hypersetup{pdfborder={0 0 0}}\href{mailto:martin.a.andersen@hiof.no}{jonathan.beaulieu.2@ens.etsmtl.ca}}}
\and 
\IEEEauthorblockN{Francis Bordeleau~{\hypersetup{pdfborder={0 0 0}}\orcidlink{0000-0001-7727-3902}}}
\IEEEauthorblockA{\textit{\'{E}cole de techonologie sup\'{e}rieure (ETS)} \\
Montreal, Canada \\
{\hypersetup{pdfborder={0 0 0}}\href{mailto:francis.bordeleau@etsmtl.ca}{francis.bordeleau@etsmtl.ca}}}
\and 
\IEEEauthorblockN{Joachim Denil~{\hypersetup{pdfborder={0 0 0}}\orcidlink{0000-0002-4926-6737}}}
\IEEEauthorblockA{
\textit{University of Antwerp} \\
Antwerp, Belgium \\
{\hypersetup{pdfborder={0 0 0}}\href{mailto:joachim.denil@uantwerpen.be}{joachim.denil@uantwerpen.be}}}
\and 
\IEEEauthorblockN{Joost Mertens\hypersetup{pdfborder={0 0 0}}\orcidlink{0000-0002-8148-5024}}
\IEEEauthorblockA{
\textit{University of Antwerp} \\
Antwerp, Belgium \\
{\hypersetup{pdfborder={0 0 0}}\href{mailto:joost.mertens@uantwerpen.be}{joost.mertens@uantwerpen.be}}}
}

\maketitle

\begin{abstract}
The new \SysML adds mechanisms for the built-in specification of domain-specific concepts and language extensions. 
This feature promises to facilitate the creation of \DSLs and interfacing with existing system descriptions and technical designs. 
In this paper, we review these features and evaluate \SysML's capabilities using concrete use cases.
We develop \DarTwinDSL, a \DSL that formalizes the existing \DarTwin notation for \DT evolution, through \SysML, thereby supposedly enabling the wide application of \DarTwin's evolution templates using any \SysML tool.
We demonstrate  \DarTwinDSL, but also point out limitations in the currently available tooling of \SysML in terms of graphical notation capabilities. 
This work contributes to the growing field of Model-Driven Engineering (MDE) for \DTs and combines it with the release of \SysML, thus integrating a systematic approach with \DT evolution management in systems engineering.
\end{abstract}


\section{Introduction}


Facilitated by frameworks such as EMF~\cite{steinbergEMFEclipseModeling2008}, Ecore, and language workbenches such as Xtext~\cite{eysholdtXtextImplementYour2010}, MPS~\cite{MPSDomainSpecificLanguage} and MetaEdit+~\cite{kellyCollaborativeModellingMetamodelling2021}, and Eclipse Sirius~\cite{viyovicSiriusRapidDevelopment2014}, the popularity of \DSLs has continuously increased. 
Regardless, users still have to provide their own visual syntax, (executable) semantics, code generators, etc.
Alternatively, approaches such as UML Profiles enable some extension and adaptation of existing modelling languages for custom purposes. 

The release of \SysML~\cite{Seidewitz_2024} \cite{SysMLv2OMG}, the successor to the popular \SysMLone language, ceases its profile-based ties with UML and positions the language as a standalone development. 
The new \SysML provides native extension points for the specification of domain-specific concepts, suggesting the creation of \DSLs that expose: 
1) a precise syntax and semantics;
2) the reuse of associated tooling; and 
3) the seamless integration of technical design (in native \SysML) and domain-specific concepts.

In this paper, we review these promises by applying \SysML's domain-specification features to a concrete example, namely the modelling of \DTSs evolution.
Concretely, we implement the \DarTwin notation~\cite{Mertens2024} and use it to model the evolution of two use cases.

We show that, based on \SysML, \DarTwinDSL enables the formalized modelling of system evolution using precise language constructs. \DarTwinDSL then enables tool-supported reasoning on a \DTS, its \DT purposes and goals, properties, and implementation. Our tool-integrated language thus supports the creation of a catalogue of stereotypical \DTS transformations that can be applied, instantiated, and extended by system developers for their own systems.
Moreover, using \SysML to define \DarTwin could enable a seamless continuation from pure \DarTwin descriptions to detailed architecture and design DT models in \SysML.

Specifically, this paper makes the following contributions. 
(a) We develop \DarTwinDSL, a formalisation of the \DarTwin notation through \SysML that enables precise modelling of \DT evolution.
By formalising the \DarTwin notation, we found various ambiguities in the evolution model, and resolved these in the \DarTwinDSL and through a systematic application of the evolution patterns.
(b) We integrate \DarTwin with \SysML by leveraging its extensibility mechanisms, making our approach compatible with standard systems engineering tools and reducing the entry barrier for adoption. (c) We validate our approach through two case studies: a gantry crane, and a strawberry cultivation system.
(d) We also critically assess current SysMLv2 tooling capabilities and limitations for implementing domain-specific graphical notations. 

The rest of this paper is structured as follows. \Cref{sec:background} provides background on \DarTwin and \SysML. Then \cref{sec:dartwindsl} gives a detailed overview of \DarTwinDSL, outlining its design principles and leveraging of \SysML's ability to extend concepts and keywords to facilitate the operationalisation of \DarTwin evolution. Afterwards, \cref{sec:casestudies}, presents two case studies in which we validate \DarTwinDSL's applicability and capacity to model diverse system changes. Next, \cref{sec:discussion} highlights the benefits and limitations encountered and critically discusses the ongoing design considerations surrounding the approach. Thereafter, \cref{sec:related-work} reviews the related work. Finally, \cref{sec:conclusion} concludes with a discussion of future research directions.

\section{Background}
\label{sec:background}

In this section, we provide the background knowledge of our approach. First, we briefly outline the \DarTwin notation, which forms the foundation of our work. Then, we discuss support for domain-specific modelling in \SysML, which enables the implementation of \DarTwin as a \DSL.

\subsection{\DarTwin notation}
\DT services, such as predictive maintenance, advanced monitoring, and model- and data-driven optimizations, require continuous flows of data and controls between an \AT, which can be a physical object, system, or process, and its digital counterpart. These services are enabled by leveraging different techniques such as \ac{mde}, data-processing techniques, formal methods, simulation, and \ac{ai}.
A Digital Twin System (\DTS) encompasses this entire ecosystem -- the \AT, its digital counterpart(s) (i.e. the DT(s)), and the bidirectional connections between them that enable monitoring, control, and optimization. Like any other system and software artefact, a \DTS is subject to permanent and continuous evolution~\cite{Mertens2024} that can affect all its aspects~\cite{tao2022digital}.
In~\cite{Mertens2024}, we classified \DTS evolution into three types~\cite{Mertens2024}:
1) Changes of the \AT or its environment,
2) Modification of the \DTs, and
3) Changes to a \DT's \emph{purpose} and/or goals. 

Hereby, Type 1) can be seen as a classical systems engineering problem, where the system degrades (\eg due to wear and tear), system components are modified/improved, or the \AT's environment changes, while 
Type 2) involves improvements, enhancements or corrections to the \DT originating from having evaluated the \DT over some time.
Type 3) typically also comes from evaluating the system and seeing opportunities that were not evident at the original design time. It can also come from new requirements or shifting priorities from business stakeholders.
In practice, an evolution may also consist of a combination of these types.

To overcome the complexity of modern systems, engineers must treat design and evolution with the required care.
Thus, system builders will rely on modelling languages and \ac{mde} techniques to plan their systems, for example using an appropriate modelling language such as Ptolemy II~\cite{ptolemaeus2014system}, Modelica~\cite{mattsson1998physical} or SysML~\cite{omg:sysml:1.5}.
To avoid invalidating the \DTS or causing damage to the \AT when facing system evolution, but also to manage the complexity of the evolution itself, the evolution should be planned, documented, and implemented systematically.
\DarTwin~\cite{Mertens2024} is a recently developed notation focusing on \DT evolution. So far, however, this notation has not been formally defined nor tool supported. 

\begin{figure}
\centering
\includegraphics[width=1.0\linewidth]{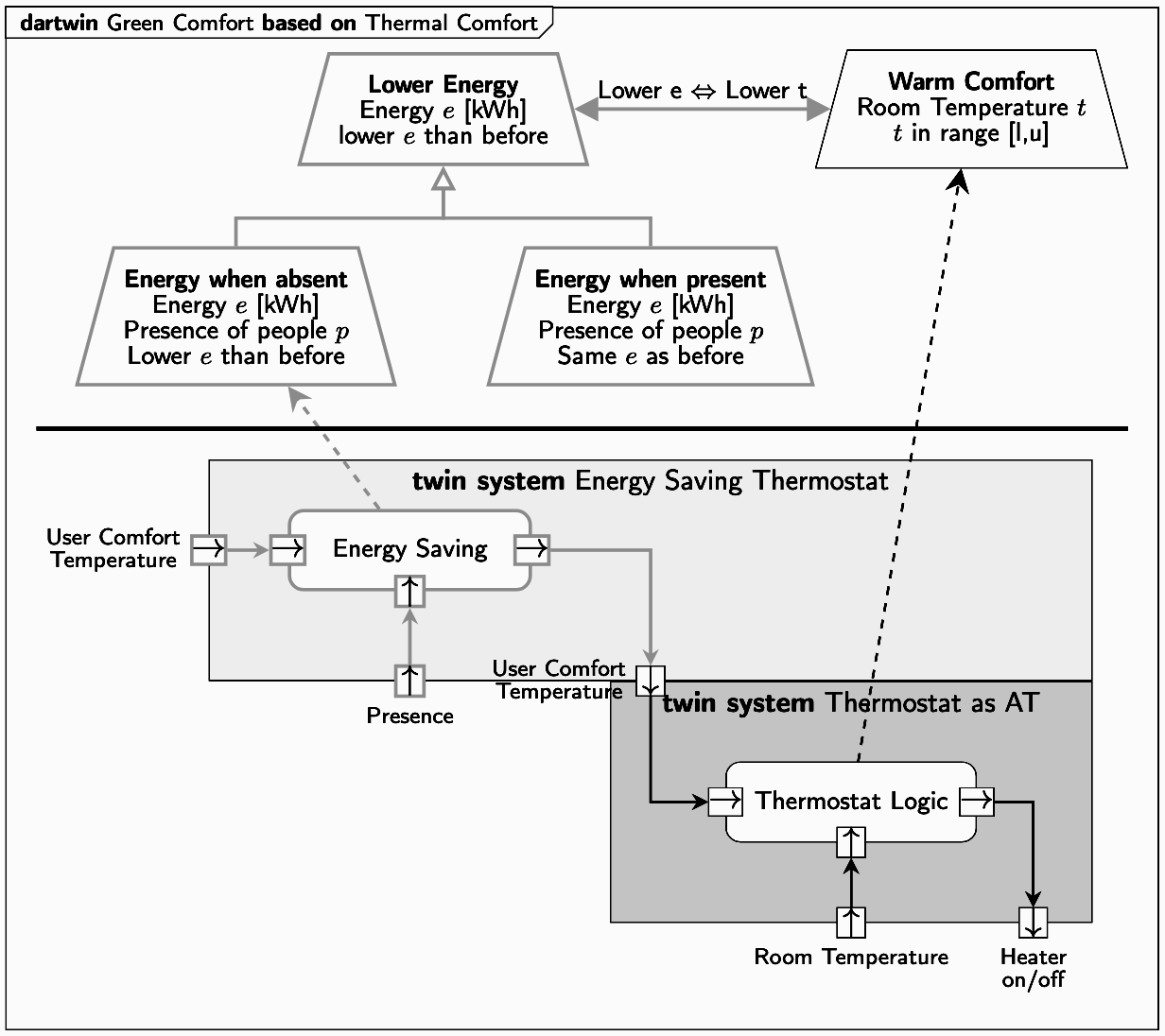}
\caption{\DarTwin notation example (from~\cite{Mertens2024}).}
\label{fig:schema-new-goal} 
\end{figure}

In \cite{Mertens2024}, we introduced \DarTwin as a graphical notation to support the systematic evolution of \DTs in the context of \DTSs. In this paper, a running example is used to illustrate how DT evolution can be described using a set of generic evolution transformations.
An example of the \DarTwin notation is shown in \Cref{fig:schema-new-goal}. This notation enables the modelling of \DT \emph{goals} and their relations, and how these goals are realized by an architecture of \DTs inside a \DTS, connected via \emph{ports}.
\DarTwin was intentionally designed to give an abstract, graphical overview of \DTs and their goals. In \DarTwin graphical view, the goals are displayed at the top, whilst the composite structure of the \DTs and their connections sit at the bottom. A horizontal bar separates the two sections.

\subsection{Domain Specific modelling in \SysML}
The release of \SysML~\cite{Seidewitz_2024} provides developers with a capable tool to design and analyse their systems. 
As a direct successor to \SysMLone~\cite{omg:sysml:1.5}, \SysML is seeing widespread interest and is predicted to become a de-facto standard for systems modelling, which further reduces the entry-barrier for \DarTwinDSL's adoption. 

Contrary to \SysMLone, the new version cuts ties with UML and streamlines its specification paradigm. 
Notably, the language is built for extensibility and enables the creation of custom keywords and concepts. This enables users to create domain-specific models with special concepts that are more intuitive for the domain experts, but still defined precisely through \SysML. This combination of domain concepts and \SysML is what we shall explore for the \DarTwin notation.

Two mechanisms are essential for creating a \DSL: specialization and redefinition. Specialization (denoted by \lstinline{:>}) allows a model element to inherit properties from another element while potentially adding new characteristics. Redefinition (denoted by \lstinline{:>>}) enables a form of inheritance where the specialized element can modify or override the characteristics of its parent. These mechanisms create the foundation for language engineering in \SysML, allowing domain-specific concepts to be formally anchored in the \SysML language's core.

\section{\DarTwin the \DSL on \SysML}
\label{sec:dartwindsl}

The \DarTwin notation paved the way for a systematic description of \DT evolution. 
Nonetheless, we point out certain limitations in the preceding work:
\begin{enumerate}
\item The notation focused on a syntactic description (and relation) of concepts related to \DT evolution without providing any precise semantics.
This limits its application to informal documentation rather than enabling systematic engineering processes. We want to explore the possibility of turning \DarTwin into a \DSL with a precise semantics.
\item \DarTwin remained a notation without proper tool support to operationalise the models. 
This makes it challenging to integrate \DarTwin into existing development workflows, verify the correctness of evolution steps, or automate any part of the evolution process, limiting its practical utility in digital twin projects. We want to explore whether there are already tools available that could provide support for our emerging \DarTwinDSL.
\item Based on the former two limitations, this means that \DarTwin is currently not actionable, meaning that we cannot implement automated reasoning and other features that require tools support.
\item The general evolution transformations of~\cite{Mertens2024} were derived in a setting of \acp{cps}; they may not carry over to other domains. Furthermore, the list of transformations is probably non-exhaustive.
\end{enumerate}

To mitigate the shortcomings of the \DarTwin notation we have implemented \DarTwin as an \emph{embedded \DSL} in \SysML. 
This should imply several advantages. 
First, it renders \DarTwin into a language with a precisely defined syntax and semantics since \SysML has mechanisms for language extension.
Second, it would mean that tools for \SysML would be applicable for \DarTwin. 
Third, by seamlessly integrating  \DarTwin with the (likely) de-facto standard in systems engineering, our approach increases the applicability of the \DarTwinDSL and facilitates its adoption. 
Fourth, we may reuse \SysML's language infrastructure, library system, model encoding, etc.

The \DarTwinDSL definition consists of a metamodel comprised by a library defined in \SysML of the \DarTwin concepts, supplemented by the declaration of the corresponding \DarTwin keywords in \SysML. \Cref{fig:xxx0} shows the conceptual metamodel of \DarTwin concepts at the bottom, the matching meta-definitions for keyword declarations (\eg \dartwinInline{\#dartwin}, 
\dartwinInline{\#twinsystem}, 
\dartwinInline{\#goal}) at the top. We provide the source code in our open-source repository\footnote{\url{https://github.com/joostm8/DarTwin-DSL/}}, an archived version is also on Zenodo \cite{zenodorepo2025}. 

\begin{figure}
\centering
\resizebox{0.7\linewidth}{!}{  

    






\begin{tikzpicture}

\node[goal] (goal1) {\textbf{Goal1}}; 

\begin{scope}[yshift=-2.25cm]  
    \node[digitaltwin,minimum width=2.5cm] (dt1) {DT1};
    \node[port,label={[inner sep=0]below left:{\tiny p12}},xshift=-2em] (dt1-in) at (dt1.south) {$\uparrow$};
    \node[port,label={[inner sep=0]below right:{\tiny p13}},xshift=2em] (dt1-out) at (dt1.south) {$\downarrow$};

   \node[port,
        label={[portlabel]below:{\tiny ts2}},
        below=.5 of dt1-in,
        ] (outer-in) {$\uparrow$};
    \node[port,
        label={[portlabel]below:{\tiny ts3}},
        below=.5 of dt1-out
        ] (outer-out) {$\downarrow$};

    \draw[flow] (outer-in.north) -- (dt1-in.south) node[portlabel,midway,right] {\tiny c2};
    \draw[flow] (dt1-out.south) -- (outer-out.north) node[portlabel,midway,left] {\tiny c3};
    
    \node[port,label={[inner sep=0]above left:{\tiny p11}}] (dt1-in2) at (dt1.west) {$\rightarrow$};
    \node[port,
        label={[portlabel]left:{\tiny ts1}},
        left=1 of dt1-in2
        ] (outer-in2) {$\rightarrow$};
    \draw[flow] (outer-in2.east) -- (dt1-in2.west) node[portlabel,midway,below] {\tiny c1};

    \node[right=.5 of dt1] (dummy-mark) {};

    \begin{scope}[on behind layer]
       \node[dtsystem,
            label={[below]north:{\textbf{twin system}} TwinSystem},
            fit={(outer-in.east)($(dt1.west)-(1em,0)$)($(dt1.north east)+(1em,1.5em)$)(outer-in2.north)(dummy-mark)}] (system) {};
    \end{scope}

\end{scope}

\draw[relationToGoal] (dt1) -- (goal1) node[midway,right,portlabel]{a1};

\node[fit=(system)(goal1),
draw,inner sep=2em,inner xsep=3.5em] (-box) {};
\node[framelabel,inner xsep=.5em,inner ysep=.25em]  (-label) at (-box.north west) {\textbf{dartwin} Basic};

\pic {frameSeparation={-box/goal1}};
\end{tikzpicture}
}
\caption{\DarTwin Basic with explicit names on ports and connections}
\label{fig:dartwin_basic} 
\end{figure}
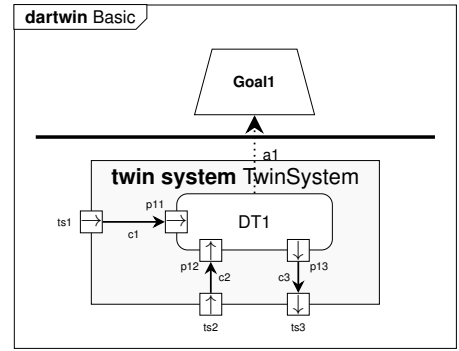

Using these domain-specific definitions, we can model \DarTwin in our textual \DSL as shown in \Cref{lst:xxx1}, which is based on the \emph{Basic} \DTS that was originally shown in~\cite{Mertens2024}. In \cf \Cref{fig:dartwin_basic} we explicitly provide the names of the ports and connections to make the correspondence to \Cref{lst:xxx1} easier. 

\begin{figure}
\centering
\resizebox{\linewidth}{!}{
\begin{tikzpicture}
\tikzset{
    every node/.append style={font=\sffamily\small,align=center},
    classpart/.append style={draw,rectangle},
    twopart/.append style={
        draw,
        rectangle split,rectangle split parts=2,
        inner sep=2pt,
    },
    metaclass/.append style={
        draw,
        rectangle split,rectangle split parts=2,
        inner sep=2pt,
    },
    twopartRound/.append style={
        draw,rounded corners=.5em,
        minimum width=10em,
        rectangle split,rectangle split parts=2,
        inner sep=.25em
    },
    threepart/.append style={
        draw,
        rectangle split,rectangle split parts=3,
        inner sep=2pt,
    },
    inherit/.append style={
    
    }
}
    
\node[metaclass] (DarTMetadata) {
\guillemotleft \textit{metadata def}\guillemotright\\$<$\textcolor{dartwin_purple}{dartwin}$>$ DarTMetadata \nodepart{two}
};

\node[metaclass,below=1 of DarTMetadata] (TSMetadata) {
\guillemotleft \textit{metadata def}\guillemotright\\$<$\textcolor{dartwin_purple}{twinsystem}$>$ TSMetadata \nodepart{two}
};

\node[metaclass,below=1 of TSMetadata] (DTMetadata) {
\guillemotleft \textit{metadata def}\guillemotright\\$<$\textcolor{dartwin_purple}{digitaltwin}$>$ DTMetadata \nodepart{two}
};

\node[metaclass,below right=0 and 1 of DarTMetadata.north east] (ConflictMetadata) {
\guillemotleft \textit{metadata def}\guillemotright\\$<$\textcolor{dartwin_purple}{vs}$>$ ConflictMetadata \nodepart{two}
};

\node[metaclass,below right=0 and 1 of TSMetadata.north east] (GoalMetadata) {
\guillemotleft \textit{metadata def}\guillemotright\\$<$\textcolor{dartwin_purple}{goal}$>$ GoalMetadata \nodepart{two}
};

\node[metaclass,below right=0 and .75 of DTMetadata.north east] (ArbiterMetadata) {
\guillemotleft \textit{metadata def}\guillemotright\\$<$\textcolor{dartwin_purple}{arbiter}$>$ ArbiterMetadata \nodepart{two}
};

\node[metaclass,below right=0 and 1 of ConflictMetadata.north east] (DarTransMetadata) {
\guillemotleft \textit{metadata def}\guillemotright\\$<$\textcolor{dartwin_purple}{dartrans}$>$ DarTransMetadata \nodepart{two}
};

\node[metaclass,below=.5 of DarTransMetadata] (BeforeMetadata) {
\guillemotleft \textit{metadata def}\guillemotright\\$<$\textcolor{dartwin_purple}{dartwin\_before}$>$ BeforeMetadata \nodepart{two}
};

\node[metaclass,below=.5 of BeforeMetadata] (CoreMetadata) {
\guillemotleft \textit{metadata def}\guillemotright\\$<$\textcolor{dartwin_purple}{dartwin\_core}$>$ CoreMetadata \nodepart{two}
};

\node[metaclass,below=.5 of CoreMetadata] (AfterMetadata) {
\guillemotleft \textit{metadata def}\guillemotright\\$<$\textcolor{dartwin_purple}{dartwin\_after}$>$ AfterMetadata \nodepart{two}
};

\node[draw,rectangle,inner sep=1em,
    fit=(DarTMetadata)(DTMetadata)(ConflictMetadata)(ArbiterMetadata)
        (BeforeMetadata)(AfterMetadata)] (DarTwinMetadata-box) {};
\node[trapezium,draw,trapezium left angle=0, trapezium right angle=75,above right=0 and 1.12 of DarTwinMetadata-box.north west] {DarTwinMetadata};


\begin{scope}[yshift=-9cm,xshift=1cm, every node/.append style={transform shape}]

\node[twopartRound] (DarTwin) {
\guillemotleft \textit{part}\guillemotright\\DarTwin \nodepart{two}
\textit{allocations}\\noname connect goals to\\DarTwin.twin\_system.digital\_twin
};

\node[twopartRound,below=1.5 of DarTwin] (core) {
\guillemotleft \textit{part}\guillemotright\\ core :$>$ DarTwin::DarTwin \nodepart{two}
};

\node[twopartRound,minimum width=5em,below left=0.75 and 0.25 of core.south] (before) {
\guillemotleft \textit{part}\guillemotright\\ before :$>$ core 
};

\node[twopartRound,minimum width=5em,right=0.5 of before] (after) {
\guillemotleft \textit{part}\guillemotright\\ afterw :$>$ core 
};

\draw[sysml-inherit] (core) -- (DarTwin);
\draw[sysml-inherit] (before) -- (core);
\draw[sysml-inherit] (after) -- (core);

\node[draw,rectangle,inner sep=1em,
    fit=(core)(before)(after)] (DarTrans-box) {};
\node[trapezium,draw,trapezium left angle=0, trapezium right angle=75,above right=0 and .5 of DarTrans-box.north west] {DarTrans};

\node[twopartRound,right=.7 of DarTwin.north east] (arbiter) {
\guillemotleft \textit{part}\guillemotright\\arbiter1: Arbiter 
};
\draw[sysml-partof] (arbiter) -- (DarTwin);

\node[twopart,right=.7 of arbiter] (arbiter-partdef) {
\guillemotleft \textit{part def}\guillemotright\\Arbiter \nodepart{two}
};
\draw[sysml-defined] (arbiter) -- (arbiter-partdef);

\node[twopartRound,minimum width=5em,right=.7 of arbiter-partdef] (input) {
\guillemotleft \textit{port}\guillemotright\\inputs[2..*] \nodepart{two}
};
\node[twopartRound,minimum width=5em,below=0.1 of input] (output) {
\guillemotleft \textit{port}\guillemotright\\output[1] \nodepart{two}
};

\draw[sysml-whitepart] (input) -- (arbiter-partdef);
\draw[sysml-whitepart] (output) -- (arbiter-partdef);

\node[twopartRound,below=0.25 of arbiter] (twinsystem) {
\guillemotleft part\guillemotright\\twin\_system \nodepart{two}
};

\node[twopartRound,below=0.25 of twinsystem] (requirement) {
\guillemotleft \textit{requirement}\guillemotright\\goals: Goal[*] \nodepart{two}
};
\draw[sysml-partof] (twinsystem) -- (DarTwin);
\draw[sysml-partof] (requirement) -- (DarTwin);

\node[twopartRound,below right=of twinsystem] (digitaltwin) {
\guillemotleft \textit{part}\guillemotright\\digital\_twin[*] \nodepart{two}
};
\draw[sysml-partof] (digitaltwin) -- (twinsystem.south east);

\node[threepart,below=.75 of requirement] (goal) {
\guillemotleft \textit{requirement def}\guillemotright\\Goal \nodepart{two}
\textit{doc}\\Our purpose  \nodepart{three}
\textit{subject}\\digital\_twin
};

\draw[sysml-defined] (requirement) -- (goal);
\draw[sysml-dotted] (requirement.east) -- (digitaltwin.west);

\node[twopart,below right=.75 and 2 of goal.north east] (conflict) {
\guillemotleft \textit{connection def}\guillemotright\\Conflict \nodepart{two}
\textit{attributes}\\explanation: String
};

\draw[sysml-whitepart] (goal.east) -- node[above,midway] {g1} ($(conflict.west)+(0,1em)$);
\draw[sysml-whitepart] ($(goal.east)-(0,2em)$) -- node[above,midway] {g2} ($(conflict.west)-(0,1em)$);

\node[draw,rectangle,inner sep=1em,
    fit=(digitaltwin)(arbiter)(goal)(output)(DarTwin)(conflict)(DarTrans-box)] (DarTwin1-box) {};
\node[trapezium,draw,trapezium left angle=0, trapezium right angle=75,above right=0 and .45 of DarTwin1-box.north west] {DarTwin};
\end{scope}


\draw[-{Stealth[scale=1]},thick,dashed] (DarTwinMetadata-box) -- node[midway,right]{\guillemotleft import\guillemotright*} (DarTwinMetadata-box.south |- DarTwin1-box.north);

\end{tikzpicture}
}
\caption{\DarTwin metamodel}
\label{fig:xxx0}
\end{figure}

\begin{remark}
When we refer to the original \DarTwin notation, we refer to a graphic form that we draw manually from graphic building blocks. This remains the syntactic form we would like our tooling through \SysML to obtain, but which turns out to be difficult to achieve. This will be further discussed later in the paper.
\end{remark}

\begin{lstlisting}[language=dartwin,float,
caption={\DarTwin Basic in textual form},
label={lst:xxx1},belowcaptionskip=0.25cm,]
#dartwin Basic {
  #twinsystem TwinSystem {
    #digitaltwin DT1 {			
      port p11;
      port p12;
      port p13;
    }			
    connection c1 connect Basic.AT.ts1 to DT1.p11;			
    connection c2 connect Basic.AT.ts2 to DT1.p12;
    connection c3 connect DT1.p13 to Basic.AT.ts3;              
  } // TwinSystem		
  part AT {
    port ts1;
    port ts2;
    port ts3;
  }
  #goal Goal1 {
    doc /* Goal 1 */
  }						
  allocation a1 allocate Goal1 to TwinSystem.DT1;
} // Basic DarTwin

\end{lstlisting}

In~\cite{Mertens2024} we used the term \DarTwin to describe a \DTS, as well as to describe the evolution from one version of the \DTS to another. When we worked on making our notation into a language, we realized that a transition is different from a single \DTS. We had already depicted the elements of the transition in orange, but we also realized that in our examples in~\cite{Mertens2024} the changes were only additions, while in the general case a transition might imply removals as well as additions.

How should we define removal of elements? \SysML does not define such self-modifying operations. We considered applying the variability mechanisms, but after some exploration decided to apply the well-known object-oriented concepts of specialization.

\paragraph*{\textbf{DarTrans Transformations}}
While the purpose of \DarTwin is to describe the evolutions of \DTSs, we note that the \DarTwin in \Cref{lst:xxx1}  defines an initial \DTS starting point. 
To define the evolution transition, we extend \DarTwin to include the concept of \emph{DarTrans} transformations.

A \dartwinInline{\#dartrans} transformation in \DarTwinDSL is a complete description of an evolution from an initial \DarTwin to an evolved \DarTwin.

\dartwinInline{\#dartrans} transformations represent three interrelated models that in combination make up the transformation.
\begin{itemize}
\item \dartwinInline{\#dartwin\_core}: all the common elements that do not change during the evolution. 
\item \dartwinInline{\#dartwin\_before} specializes \dartwinInline{\#dartwin\_core} and lists the elements that will be \textit{deleted} in the evolution.
Together with the inherited elements from \dartwinInline{\#dartwin\_core} they represent the evolution starting point.
\item \dartwinInline{\#dartwin\_after} specializes \dartwinInline{\#dartwin\_core} and lists the \textit{added} elements of the evolution.

Note that \textit{changing} or \textit{updating} an element is modeled as a combination of deletion of the old, and updating of the new version. Thus, a modified element should be present in both \dartwinInline{\#dartwin\_before} and \dartwinInline{\#dartwin\_after}.
Together with the inherited elements from \dartwinInline{\#dartwin\_core} it represents the result of the evolution.
\end{itemize}

Visually, a \dartwinInline{\#dartrans} is depicted similarly to a \DarTwin model, with the difference that changes in components and connections are drawn in orange colour.
Specifically, \dartwinInline{\#dartwin\_before} shows  deleted elements using dashed lines, while elements that are updated or added by \dartwinInline{\#dartwin\_after} are solid.

\Cref{lst:xxx3} shows how the \texttt{OrthogonalWithNewOutput} evolution is defined with these three models. As this transformation only defines additions, \texttt{OrthogonalWithNewOutput\_before} is equivalent to \texttt{OrthogonalWithNewOutput\_core} (\cf Line 3).

\begin{lstlisting}[language=dartwin,float,
caption={\DarTwin evolution OrthogonalWithNewOutput},
label={lst:xxx3}]
#dartrans OrthogonalWithNewOutput {
	#dartwin_core OrthogonalWithNewOutput_core :> Basic;
	#dartwin_before OrthogonalWithNewOutput_before :> OrthogonalWithNewOutput_core;
	#dartwin_after OrthogonalWithNewOutput_after :> Basic{
		#twinsystem :>> TwinSystem {
			#digitaltwin DT2 {
				port p21;
				port p22;
			}
			connection c4 connect OrthogonalWithNewOutput.OrthogonalWithNewOutput_after.AT.ts1 to DT2.p21; 
			connection c5 connect DT2.p22 to OrthogonalWithNewOutput.OrthogonalWithNewOutput_after.AT.ts4;
		}
		part :>> AT {
			port ts4;
		}
		#goal Goal2 {
			doc /* Goal 2 */
		}			
		allocation a2 allocate Goal2 to TwinSystem.DT2;
	}
}	

\end{lstlisting}

The graphic rendering of \texttt{OrthogonalWithNewOutput} is shown in \Cref{fig:xxx4}. Orange colour indicates added elements, \ie those described in \texttt{OrthogonalWithNewOutput\_after} (\cf in \Cref{lst:xxx3}, Lines 4 -- 20).

\begin{figure}
\centering
\resizebox{.65\linewidth}{!}{
\begin{tikzpicture}
\node[goal] (goal1) {\textbf{Goal1}}; 
\node[goal,highlight,right=.9 of goal1] (goal2) {\textbf{Goal2}};

\begin{scope}[yshift=-2.25cm]
    \node[digitaltwin,minimum width=2cm] (dt1) {DT1};
    \node[port,label={below:{}},xshift=-1em] (dt1-in) at (dt1.south) {$\uparrow$};
    \node[port,label={below:{}},xshift=1em] (dt1-out) at (dt1.south) {$\downarrow$};

    \node[digitaltwin,highlight,minimum width=2cm,right=.5 of dt1] (dt2) {DT2};
    \node[port,highlight,label={below:{}},xshift=-1em] (dt2-in) at (dt2.south) {$\uparrow$};
    \node[port,highlight,label={below:{}},xshift=1em] (dt2-out) at (dt2.south) {$\downarrow$};

    \node[port,
        label={[portlabel]below:{}},
        below right=.5 and 1 of dt1-in,
        ] (outer-in) {$\uparrow$};
    \node[port,
        label={[portlabel]below:{}},
        below right=.5 and 1 of dt1-out
        ] (outer-out) {$\downarrow$};

    \node[port,highlight,
        label={[portlabel]below:{}},
        below=.5 of dt2-out
        ] (outer-out2) {$\downarrow$};

    \draw[flow] (outer-in.north) -- (dt1-in.south);
    \draw[flow] (dt1-out.south) -- (outer-out.north);
    
    \draw[flow,highlight] (outer-in.north) -- (dt2-in.south);
    \draw[flow,highlight] (dt2-out.south) -- (outer-out2.north);

    \node[port,label={below:{}}] (dt1-in2) at (dt1.west) {$\rightarrow$};
    \node[port,
        label={[portlabel]below:{}},
        left=.35 of dt1-in2
        ] (outer-in2) {$\rightarrow$};
    \draw[flow] (outer-in2.east) -- (dt1-in2.west);

    \begin{scope}[on behind layer]
       \node[dtsystem,
            label={[below]north:{\textbf{twin system}}},
            fit={(outer-in.east)($(dt1.west)-(1em,0)$)($(dt2.north east)+(1em,1.5em)$)(outer-in2.north)}] (system) {};
    \end{scope}
    
\end{scope}

\draw[relationToGoal] (dt1) -- (goal1);
\draw[relationToGoal,highlight] (dt2) -- (goal2);

\pic (fg-frame) {dartransFrameFit={OrthogonalWithNewOutput/Basic/(goal1)(system)(outer-in)}};
\pic {frameSeparation={fg-frame-box/goal1}};
\end{tikzpicture}
}
\caption{DarTrans evolution \texttt{OrthogonalWithNewOutput}}
\label{fig:xxx4}
\end{figure}

\section{\DarTwin tooling with \SysML tooling?}
\label{sec:casestudies}

As a means of checking the completeness and applicability of the \DarTwinDSL, we look at two case studies that feature evolution: a strawberry cultivation system and the gantry crane system~\cite{Mertens2024}. Their implementations can also be found in our repository.

\subsection{Strawberry Cultivation System as a Foundational Example}\label{sec:Strawberry}
As an initial test of \DarTwinDSL's applicability within \SysML, we modeled a moderately complex system: a \DTS for controlled indoor strawberry cultivation.

\begin{figure}
\centering

\resizebox{.8\linewidth}{!}{%
\begin{tikzpicture}

\node[goal] (increase-yield-goal) {\textbf{Increase Yield}\\yield $y$\\
        higher $y$ than before};
\node[goal,right=of increase-yield-goal] (decrease-water-goal) {\textbf{Decrease Water}\\water consumption $w$\\
        lower $w$ than before};

\begin{scope}[xshift=2cm,yshift=-3cm]  

    \node[digitaltwin,minimum width=4cm,minimum height=2cm] (strawberry-dt-dt) {Strawberry DT};

    \node[port,label={below:{}}] (-irrigation) at ($(strawberry-dt-dt.south)!.75!(strawberry-dt-dt.south west)$) {$\downarrow$};
    \node[port,label={below:{}}] (-human) at (strawberry-dt-dt.south) {$\downarrow$};
    \node[port,label={below:{}}] (-ventilation) at ($(strawberry-dt-dt.south)!.75!(strawberry-dt-dt.south east)$) {$\downarrow$};  

    \node[port,label={[portlabel]below:{Irrigation\\Actuator}},below=2em of -irrigation] (sys-irrigation) {$\downarrow$};
    \node[port,label={[portlabel]below:{Human\\Actuator}},below=2em of -human] (sys-human) {$\downarrow$};
    \node[port,label={[portlabel]below:{Ventilation\\Actuator}},below=2em of -ventilation] (sys-ventilation) {$\downarrow$};

    \draw[flow] (-irrigation) -- (sys-irrigation);
    \draw[flow] (-human) -- (sys-human);
    \draw[flow] (-ventilation) -- (sys-ventilation);

    \node[port,label={below:{}}] (-input) at ($(strawberry-dt-dt.east)+(0,0)$) {$\leftarrow$};
    \node[port,label={[portlabel]right:{Multi-\\sensor}},right=2em of -input] (sys-input) {$\leftarrow$};
    \draw[flow] (sys-input) -- (-input);

    \begin{scope}[on behind layer]
       \node[dtsystem,
            label={[below]north:{\textbf{twin system} \emph{Strawberry}}},
            fit={(sys-ventilation.east)($(strawberry-dt-dt.west)-(1em,0)$)($(strawberry-dt-dt.north east)+(1em,1.5em)$)
            (sys-input.north)
            }] (system) {};
    \end{scope}
    
\end{scope}

\draw[relationToGoal] (strawberry-dt-dt) -- (increase-yield-goal);
\draw[relationToGoal] (strawberry-dt-dt) -- (decrease-water-goal);

\node[fit=(system)(increase-yield-goal)(decrease-water-goal)(sys-ventilation)(sys-input),
draw,inner sep=2em,inner xsep=5.5em,xshift=1em] (-box) {};
\node[framelabel,inner xsep=.5em,inner ysep=.25em]  (-label) at (-box.north west) {\textbf{dartwin} StrawberryCultivationTrans};

\pic {frameSeparation={-box/increase-yield-goal}};
\end{tikzpicture}%
}%
\caption{Strawberry Cultivation System \DarTwin}
\label{fig:strawberry-dartwin}
\end{figure}
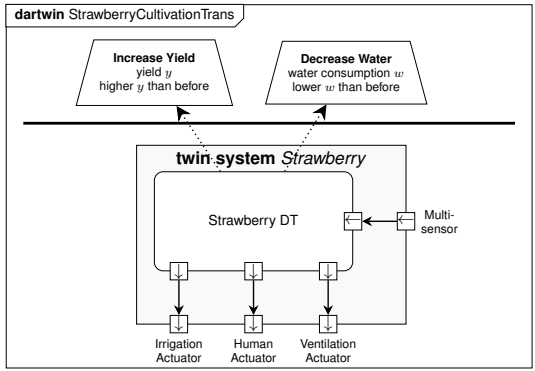

The \DT performs three primary functions: 1) monitoring environmental parameters via multisensor arrays, 2) controlling irrigation and ventilation through actuators, and 3) supporting human operators by surfacing alerts and actionable data. These responsibilities are represented by explicitly defined goals and realized through a composite twin structure as shown in the manually edited \DarTwin~\Cref{fig:strawberry-dartwin}.

\begin{figure}
    \centering
    \includegraphics[width=0.5\textwidth]{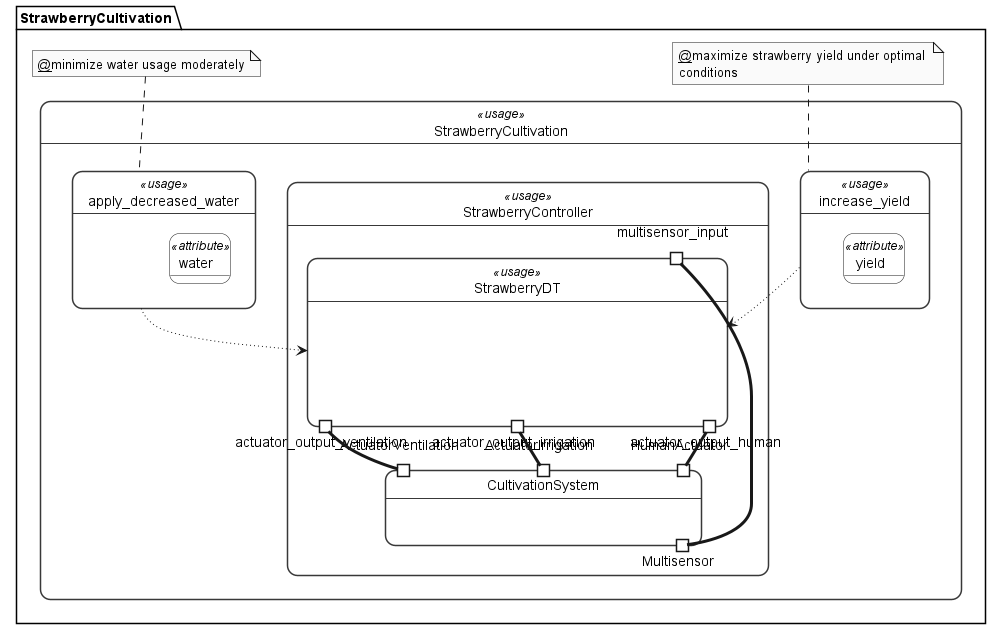}
    \caption{This rendering was produced using the pilot implementation.}
    \label{fig:strawberry-dartwin-pilot}
\end{figure}

\begin{figure}
    \centering
    \includegraphics[width=0.5\textwidth]{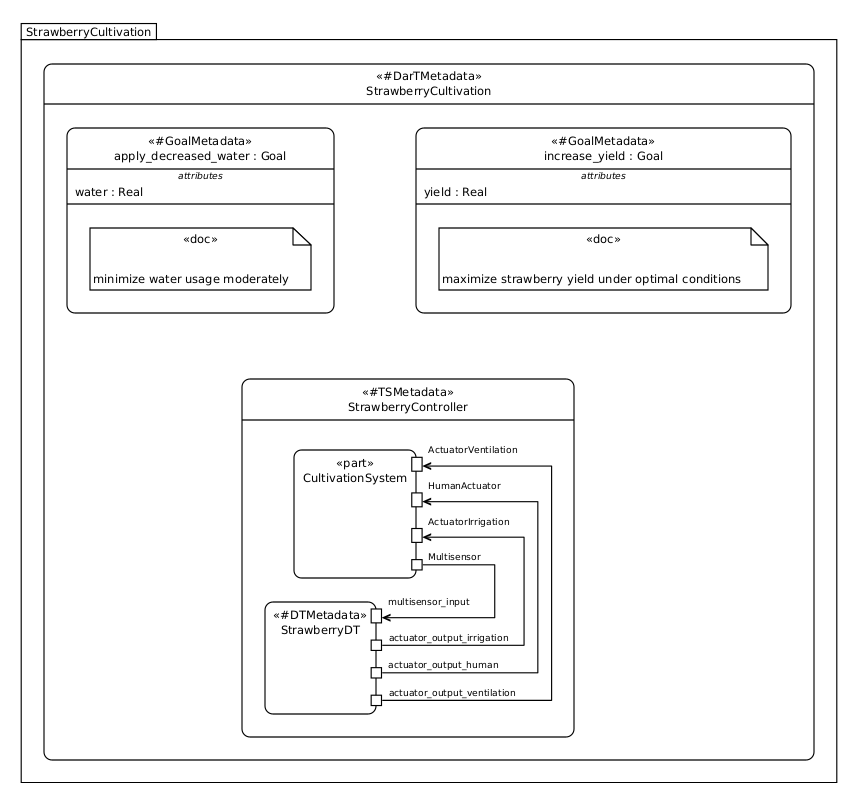}
    \caption{This rendering was produced using Tom Sawyer SysML Viewer v1.1.1 with manual adjustments to maintain DarTwin's visual style.}
    \label{fig:strawberry-dartwin-TS}
\end{figure}

\begin{figure}
\centering
\includegraphics[width=0.7\linewidth]{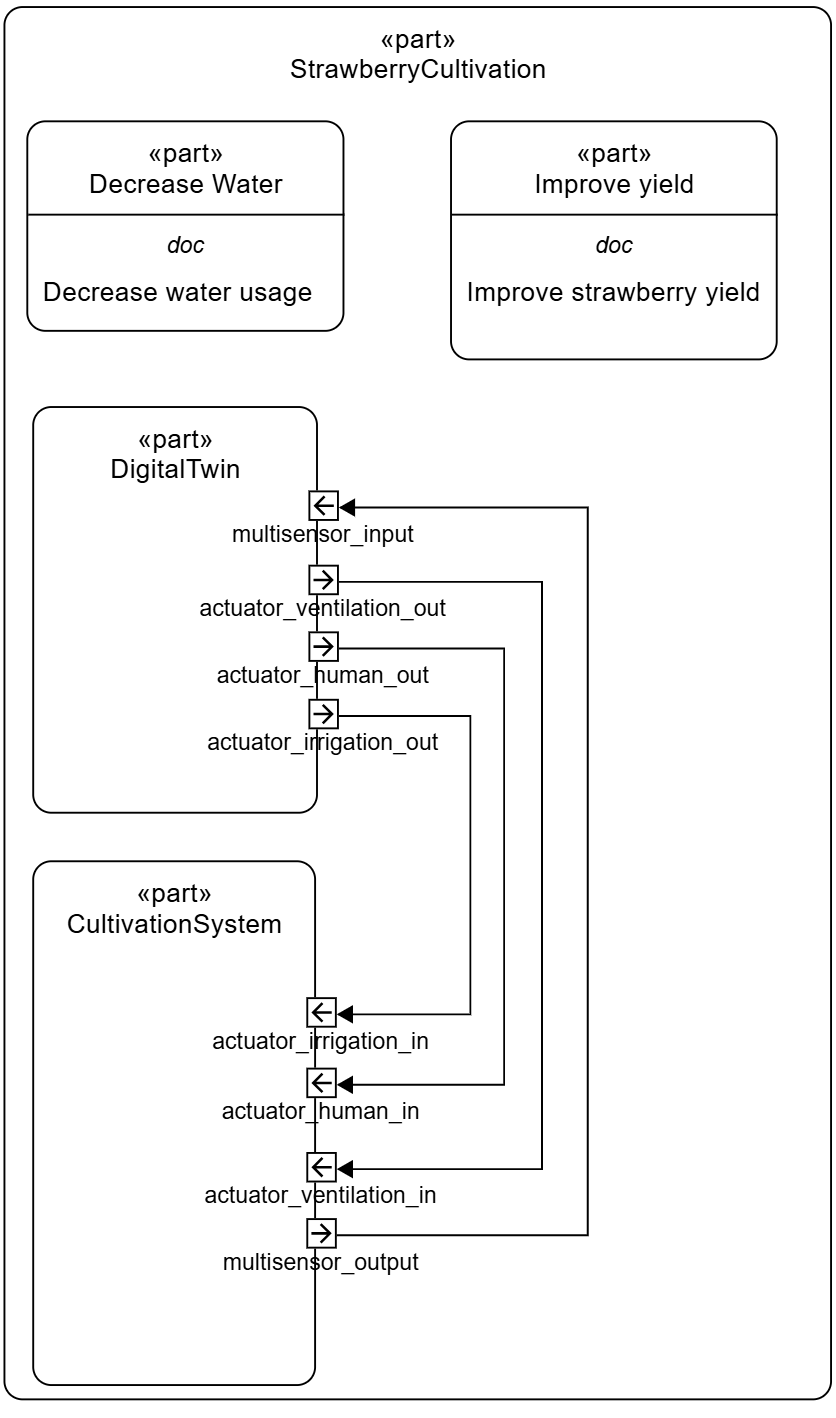}%
\caption{This rendering was produced using SysON 2025 V6.0 with manual visual style.}
\label{fig:syson}
\end{figure}

In \DarTwinDSL, the system was defined using a single \dartwinInline{\#dartwin} block containing one \dartwinInline{\#twinsystem}, multiple \dartwinInline{\#digitaltwin} elements, connections via \dartwinInline{connect}, and allocated goals through \dartwinInline{allocate}. \Cref{fig:strawberry-dartwin-pilot} provides a graphical rendering based on this model in the Pilot Implementation, while \Cref{fig:strawberry-dartwin-TS} is made through Tom Sawyer and \Cref{fig:syson} in SysON.

This use case served two purposes. First, it validated the expressive adequacy of the \DarTwinDSL. Second, it allowed assessing the capabilities and limitations of current \SysML tools for rendering such models similar to the original visual conventions proposed in~\cite{Mertens2024}.

Our findings revealed that while all tools we tested accepted the textual model, none were capable of generating graphical layouts that aligned with the intended \DarTwin visual notation. In particular:

\begin{itemize}
    \item \textbf{SysML v2 Pilot Implementation (Eclipse Plugin)}: The tool successfully parsed and executed the textual model, but the visual rendering was severely limited. Diagrams relied on a fixed PlantUML backend that cannot be modified or controlled by users. Layouts suffered from excessive whitespace and poor element placement and connection path; \lstinline{Doc} blocks rendered outside their associated \lstinline{#goal} containers.
    
    \item \textbf{Tom Sawyer SysML Viewer (v1.1.1)}: The auto-layout feature produced non-deterministic results across refreshes. Manual cleanup was necessary to remove extraneous metadata, and there was no way to enforce DarTwin’s intended visual conventions, such as top-level goal placement or horizontal placement of twin components.

    \item \textbf{SysON (v2025.6.0)}: Although the tool accepted the textual input, its interconnection diagram failed to render goals, requirements, or complex connect statements correctly—even when all port paths were fully qualified. Rendered diagrams contained anonymous parts and omitted valid elements without generating warnings or error feedback. The layout was inconsistent, and the system hierarchy was visually incoherent. The tool did not recognize language extensions such as \lstinline{#goal}, rendering the graphical view unusable for DarTwin-compliant modelling.
\end{itemize}

Despite these limitations, it is the authors' experience that the Strawberry Cultivation System demonstrates the suitability of the \DarTwinDSL metamodel for capturing structural relationships and goal allocations within \DTSs.

\subsection{Gantry Crane System}\label{sec:Gantry}

\begin{figure}
    \centering
    \includegraphics[width=0.75\linewidth]{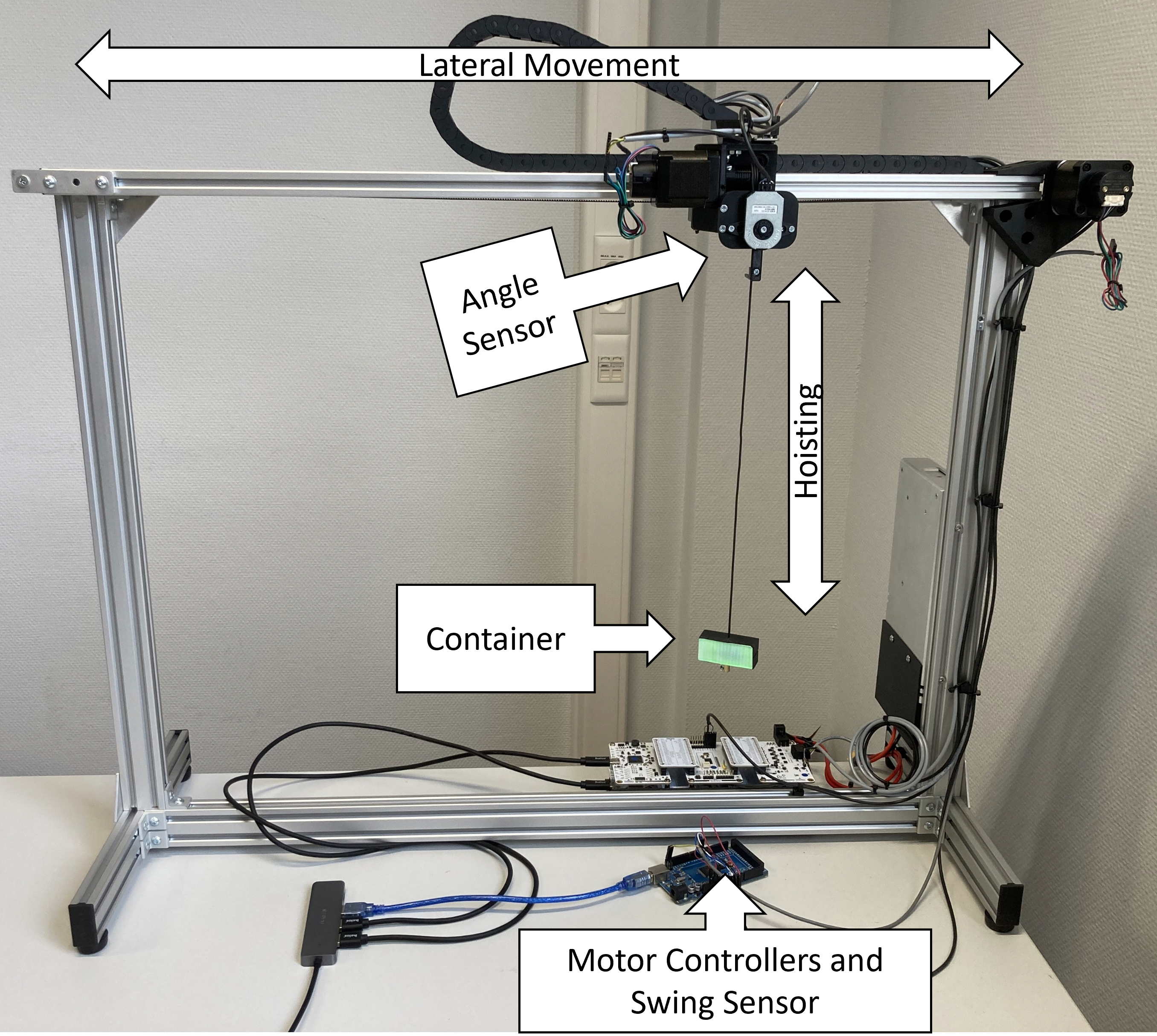}
    \caption{The 1:10th scaled gantry crane.}
    \label{fig:crane-irl}
\end{figure}

At the Univeristy of Antwerp's Cosys-Lab, a lab-scale (approximately scaled 1:10) gantry crane case study was developed to research \DTSs. The crane is depicted in \cref{fig:crane-irl}. It was inspired by the harbour of Antwerp, where such a crane moves containers from/onto docked ships. All the implementation details can be found in \cite{mertens2025labscalegantrycranedigital}. Furthermore, in \cite{Mertens2024}, evolutions of this case study were conceived to demonstrate the evolution transformations in the development of a crane \DTS. Here, we revisit one such transformation to demonstrate \dartwinInline{\#dartrans}.

\begin{figure}
\centering
\resizebox{.75\linewidth}{!}{
\begin{tikzpicture}

\pic {constraints goal};
\pic [below left=0.25 and 1 of constraints goal] {noswing goal};
\pic [below right=0.25 and 1 of constraints goal] {minimize goal};

\begin{scope}[yshift=-4.5cm]  

    \node[digitaltwin,minimum width=4cm] (trajectory-dt-dt) {Trajectory};
    \node[port,label={below:{}}] (-position) at ($(trajectory-dt-dt.south)!.75!(trajectory-dt-dt.south west)$) {$\uparrow$};
    \node[port,label={below:{}}] (-swing) at (trajectory-dt-dt.south) {$\uparrow$};
    \node[port,label={below:{}}] (-controllers) at ($(trajectory-dt-dt.south)!.75!(trajectory-dt-dt.south east)$) {$\downarrow$};  
    

    \node[port,label={[portlabel]below:{Motor\\Position}},below=2em of -position] (sys-position) {$\uparrow$};
    \node[port,label={[portlabel]below:{Swing\\Angle}},below=2em of -swing] (sys-swing) {$\uparrow$};
    \node[port,label={[portlabel]below:{Motor\\Controllers}},below=2em of -controllers] (sys-controllers) {$\downarrow$};  
    \draw[flow] (sys-position) -- (-position);
    \draw[flow] (sys-swing) -- (-swing);
    \draw[flow] (-controllers) -- (sys-controllers);
        
    \begin{scope}[on behind layer]
       \node[dtsystem,
            label={[below]north:{\textbf{twin system} \emph{Gantry Crane}}},
            fit={(sys-position.east)($(trajectory-dt-dt.west)-(1em,0)$)($(trajectory-dt-dt.north east)+(1em,1.5em)$)}] (system) {};
    \end{scope}
\end{scope}

\draw[relationToGoal] (trajectory-dt-dt) -- (constraints goal);
\draw[relationToGoal] (trajectory-dt-dt) -- (noswing goal);
\draw[relationToGoal] (trajectory-dt-dt) -- (minimize goal);

\pic (fg-frame) {dartwinFrameFit={Optimal Control//{(system)(constraints goal)(noswing goal)(minimize goal)(sys-controllers)}}};
\pic {frameSeparation={fg-frame-box/minimize goal}};
\end{tikzpicture}
}
\caption{Optimal Control \DarTwin of the Gantry Crane system}
\label{fig:gantrycrane}
\end{figure}
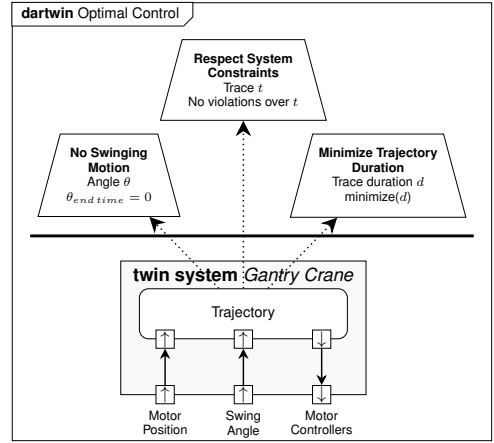

To systematically evolve a target system using a \dartwinInline{\#dartrans} evolution template, we follow the ``\emph{5-step procedure}''. We illustrate this procedure with an example from the gantry crane system. In it, we look to upgrade the \DT responsible for generating trajectories by changing it from a Linear Quadratic Regulator (LQR) to an Optimal Control Problem (OCP) solver. To do so, we apply the replacement transformation template, It can be found in \Cref{lst:5stepReplacement}. This transformation replaces one \DT by another one and reallocates the goal of the old \DT to the new one. In it, we see the three models that make up the transformation. The \dartwinInline{\#dartwin\_core} consists of the AT and its ports, as well as the goal, since these are invariant in the transformation. In the \dartwinInline{\#dartwin\_before}, we observe the original DT and its ports, connections and goal allocation. In the \dartwinInline{\#dartwin\_after}, we observe the new DT and its ports, connections and goal allocations.

\begin{lstlisting}[language=dartwin, caption={Replacement Transformation.}, label={lst:5stepReplacement}]
#dartrans Replacement{
		#dartwin_core dt_core{
			#twinsystem TS{	
			}
			part AT {
				port p1;
				port p2;
			}
			#goal goal1;
		}
		#dartwin_before dt_before :> dt_core{
			#twinsystem :>>TS{
				#digitaltwin DT1{
					port p1;
					port p2;
				} 
			connection c1 connect DT1.p1 to Replacement.dt_core.AT.p1;
			connection c2 connect Replacement.dt_core.AT.p2 to DT1.p2;
			}
			allocation a1 allocate goal1 to TS.DT1;
		} 
		#dartwin_after dt_after :> dt_core{
			#twinsystem :>>TS{
				#digitaltwin DT2{
					port p1;
					port p2;
				} 
			connection c1 connect DT2.p1 to Replacement.dt_core.AT.p1;
			connection c2 connect Replacement.dt_core.AT.p2 to DT2.p2;	
			}		
			allocation a1 allocate goal1 to TS.DT2;
		}	
	}
\end{lstlisting}

We apply this transformation to the gantry crane example system in \Cref{lst:5stepExampleSys} found below. Note that the example is in fact more complex, and is shown in its entirety in \Cref{fig:gantrycrane}, but to keep the code listings succinct, we reduced the system to only one goal and two ports instead of three of each. 

\begin{lstlisting}[language=dartwin, caption={Example system.}, label={lst:5stepExampleSys}]
#dartwin OptimalControl {
    #twinsystem GantryCrane {
        #digitaltwin TrajectoryLQR{
            port sense;
            port actuate;
        }
    }
    part PhysCrane{
        port actuate;
        port sense;
    }
    #goal NoSwing;
    connection actuation connect GantryCrane.TrajectoryLQR.actuate to PhysCrane.actuate;
    connection sensing connect PhysCrane.sense to GantryCrane.TrajectoryLQR.sense;
    allocation noSwinging allocate NoSwing to GantryCrane.TrajectoryLQR;
}	
\end{lstlisting}

In what follows, we now apply the 5-step procedure.

\begin{enumerate}
    \item We take the gantry system from \Cref{lst:5stepExampleSys} as is as input.
    \item We take the replacement pattern, and have our example specialize all the elements in its \dartwinInline{\#dartwin\_before}. This defines the premise: \emph{Can the chosen transformation pattern be applied?} This is shown in \Cref{lst:5step2}.
\begin{lstlisting}[language=dartwin,xleftmargin=-\leftmarginii,caption={Specializing  \dartwinInline{\#dartwin\_before} in the example.}, label={lst:5step2}]
#dartwin OptimalControl :> Replacement.dt_before{
    #twinsystem GantryCrane :> TS{
        #digitaltwin TrajectoryLQR :> DT1{
            port sense :> p1;
            port actuate :> p2;
        }
    }
    part PhysCrane :> Replacement.dt_before.AT{
        port actuate :> p1;
        port sense :> p2;
    }
    #goal NoSwing :> Replacement.dt_before.goal1;
    connection actuation :> Replacement.dt_before.c1 connect GantryCrane.TrajectoryLQR.actuate to PhysCrane.actuate;
    connection sensing :> Replacement.dt_before.c2 connect PhysCrane.sense to GantryCrane.TrajectoryLQR.sense; 
    allocation noSwinging :> Replacement.dt_before.a1 allocate NoSwing to GantryCrane.TrajectoryLQR;
}
\end{lstlisting}
\item We delete all elements that are part of the \dartwinInline{\#dartwin\_before} such that only elements of the \dartwinInline{\#dartwin\_core} remain. This is shown in \cref{lst:5step3}.
\begin{lstlisting}[language=dartwin,xleftmargin=-\leftmarginii,caption={Reduction to  \dartwinInline{\#dartwin\_core}.}, label={lst:5step3}]
#dartwin OptimalControl :> Replacement.dt_core{
    part GantryCrane :> TS{
    } 
    part PhysCrane :> AT{
        port actuate :> p1;
        port sense :> p2;
    }
    #goal NoSwing :> Replacement.dt_core.goal1;	
}
\end{lstlisting}
\item We add all the elements of the \dartwinInline{\#dartwin\_after} to the example system by specializing it instead of \dartwinInline{\#dartwin\_core}. The additions of the \dartwinInline{\#dartwin\_after} of the pattern may also be themselves specialized to add system-specific properties.  This is indicated in \Cref{lst:5step4}.
\begin{lstlisting}[language=dartwin,xleftmargin=-\leftmarginii,caption={Additions by specializing \dartwinInline{\#dartwin\_after}.}, label={lst:5step4}]
#dartwin OptimalControl :> Replacement.dt_after {
	#twinsystem GantryCrane :> TS{
		#digitaltwin TrajectoryOCP :> DT2{
			port sense :> p1;
			port actuate :> p2; 
		}
	}
	part PhysCrane :> Replacement.dt_after.AT{
		port actuate :> p1;
		port sense :> p2;
	}
	#goal NoSwing :> Replacement.dt_after.goal1;
	connection actuation :> Replacement.dt_after.c1 connect GantryCrane.TrajectoryOCP.actuate to PhysCrane.actuate;
	connection sensing :> Replacement.dt_after.c2 connect PhysCrane.sense to GantryCrane.TrajectoryOCP.sense;
	allocation noSwinging :> Replacement.dt_after.a1 allocate NoSwing to GantryCrane.TrajectoryOCP;
	}	
\end{lstlisting}
\item We want to merge the inherited  \dartwinInline{\#dartwin\_after} of the pattern with the transformed system. Since the pattern is very general, we may in fact just remove all references (inheritances/specializations) to the pattern to finalize the transformation. This is shown in \cref{lst:5step5}.
\begin{lstlisting}[language=dartwin,xleftmargin=-\leftmarginii,caption={Finalizing the transformation.}, label={lst:5step5}]	
#dartwin OptimalControl {
	#twinsystem GantryCrane {
		#digitaltwin TrajectoryOCP{
			port sense;
			port actuate;
		}
	}
	part PhysCrane{
		port actuate;
		port sense;
	}
	#goal NoSwing;
	connection actuation connect GantryCrane.TrajectoryOCP.actuate to PhysCrane.actuate;
	connection sensing connect PhysCrane.actuate to GantryCrane.TrajectoryOCP.sense;
	allocation noSwinging allocate NoSwing to GantryCrane.TrajectoryOCP;
}	
\end{lstlisting}
\end{enumerate}

\begin{figure}
    \centering
    \includegraphics[width=\linewidth]{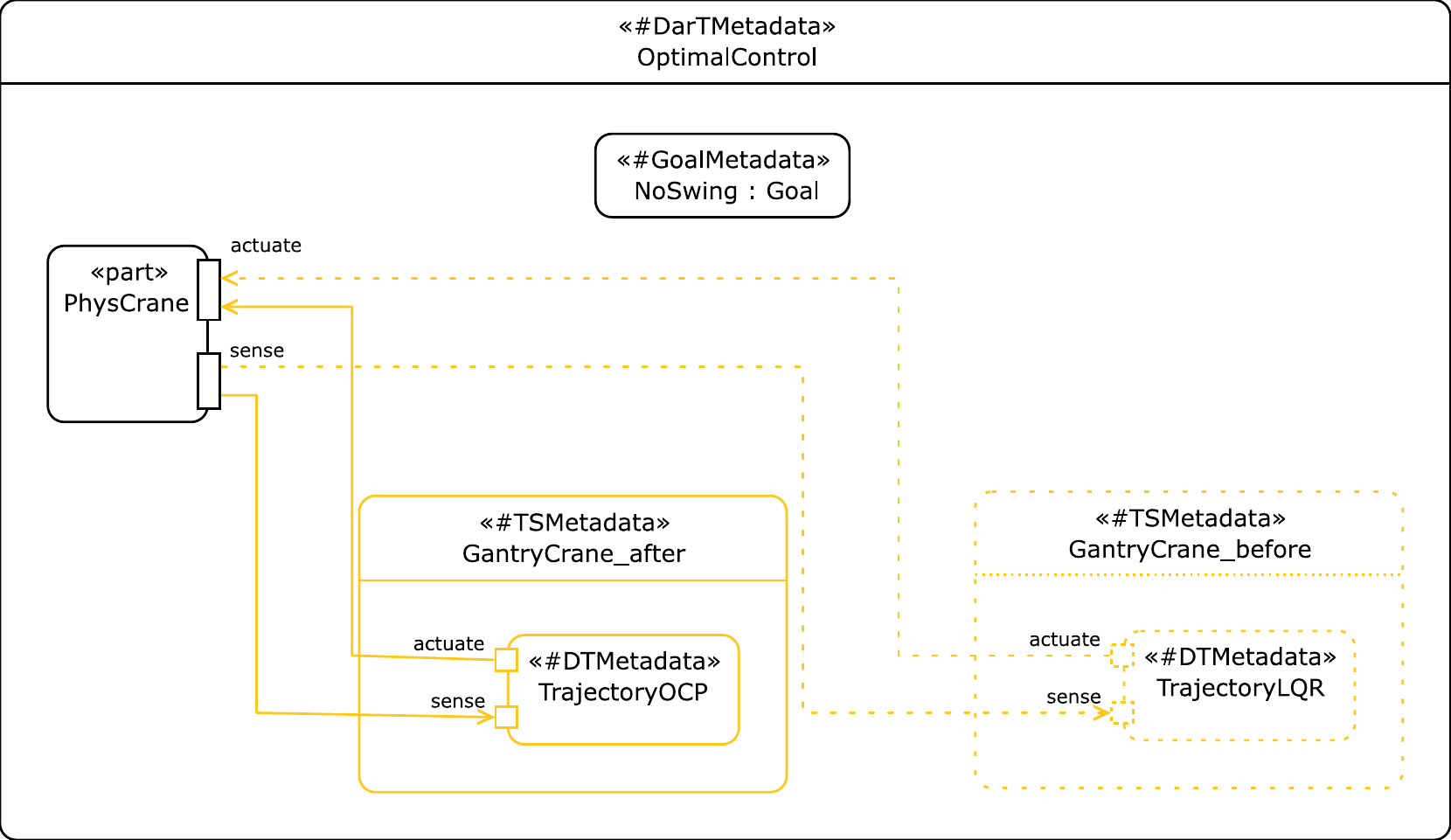}
    \caption{Visualization of the replacement transformation on the example made with the Tom Sawyer tool. Elements are highlighted as follows: deletions are dashed orange, additions are solid orange, and unchanged elements are solid black.}
    \label{fig:150pct}
\end{figure}

After following these steps, the replacement transformation is complete, and the optimal control \dartwinInline{\#dartwin} has been successfully updated. Finally, to visually show what has happened, \cref{fig:150pct} shows an overview of the transformation on the example, highlighting the changed (orange) and unchanged (black) elements. The visualization was made by exporting to SVG from the Tom Sawyer tool, which is why certain connections, e.g. the one from the goal to the digital twins, are missing. Colors were added afterward with an SVG editor (Inkscape in this case).

In the code repository, more transformation examples are listed. The main finding from applying them, is that in the original publication \cite{Mertens2024}, the transformations were interpreted rather freely. Formalizing the transformations in \DarTwinDSL made that clear, and also makes clear what is needed to correctly define a transformation.

\section{Discussion}
\label{sec:discussion}

\subsection{How to use \DarTwinDSL?}\label{sec:HowToUse}

There are two distinct ways to apply \DarTwinDSL. The first one is to define the overall design of a \DTS and its context similar to ~\cite{Mertens2024} as shown in (\cf \Cref{sec:Strawberry}) defining a \dartwinInline{\#dartwin}. The second way is to plan an evolution by deciding to apply an existing \DarTwin evolution transition patterns, as demonstrated in \Cref{sec:Gantry}.

In our experiments with this, we gathered some experiences:

\begin{itemize}
    \item A dedicated evolution tool would have been very welcome and should be possible to make.
    \item In the textual notation, keeping track of all the connections and associated ports was hard, and errors were more easily found in the graphic rendering even though the renderings were not according to the \DarTwin notation.
    \item The systematic approach to evolution was helpful, and it did reveal that some of the evolution transition patterns of \cite{Mertens2024} were not as general as they should have been. They were still useful.
\end{itemize}

\subsection{What is gained by using \SysML to define \DarTwin?}
Applying \SysML to define \DarTwin made its definition more precise. We discovered, for instance, that deletion had not been properly covered in the original \DarTwin notation.

Using the language extension mechanisms of \SysML, the tooling for \SysML should be directly applicable to handling \DarTwin. In our case, however, this was only partly true. Not all evaluated tools could handle the language extension, and we also encountered graphical rendering issues, as explained in \Cref{sec:tooling}. In addition to being unable to reproduce the view to render \DarTwin as the original notation, the tools had numerous shortcomings that made them difficult to apply.

Since \SysML has a formal definition, so will \DarTwinDSL since we define it by \SysML. This is conceptually advantageous and means that supplementary tooling can be created to support the \DarTwin method of applying evolutionary patterns as explained above.

\DarTwin's integration with \SysML also implies that any \DarTwin description can be enhanced by standard \SysML constructs. In our case, this means that the detailed design of the \DTs (e.g. state machines) can be included. In the future, we plan to support a seamless development starting from \DarTwin and ending in simulation and executable implementation.
By embedding \DarTwin in \SysML we further enable the translation of all \DarTwin concepts to standard \SysML for interoperability with other tools.

\subsection{Formalizing \DarTwin}

The primary goal behind \DarTwin's formalization was to add precision to the notation. 
By integrating it with \SysML, it should also be more applicable by extending the potential user group. While formal means make descriptions more precise, they also may make the descriptions less intuitive. The \DarTwin notation was intended to be useful for a variety of users of different backgrounds. It is not obvious that \SysML has that same effect. With the fundamental shortcomings of the available renderings of textual \SysML into graphics in the available tools we tested, we had to apply manual post-editing to make diagrams with some similarity to the original \DarTwin notation. Refer to \Cref{sec:tooling} for more on the tooling issues.

Since \SysML has included mechanisms to describe views and viewports, it is reasonable to expect that \SysML tools in the future will provide mechanisms to specify a diagram view within \SysML such that the \DarTwin notation will be produced.

There are still different ways to formalize \DarTwin also adhering to applying \SysML. Rather than applying user-defined keywords, we could have just applied the metamodel as a library of \SysML concepts. Several libraries are available with \SysML already. Our choice to apply user-defined keywords to constitute the \DarTwinDSL was motivated by considering keywords a more visible syntactic form than libraries. A set of distinguishable keywords implies that there is an integrated DSL with its own clear language definition.

We also envision \SysML libraries for the purpose of establishing a useful set of evolution transition patterns that can be applied systematically in \DT evolutions.

Our formalization makes use of existing \SysML language constructs where we do not want to make our own more specific ones. Our examples show many ports on \DTs and \AT, but a port is not a \DarTwin keyword concept - yet. Following more experience with using \DarTwin on cases, we foresee that the \DarTwin metamodel and specialized concepts will grow slightly. The compatibility of \DarTwin is not affected since the keyword definition makes it possible to go from \DarTwin keywords to equivalent \SysML constructs.

The user-defined keywords may also facilitate the making of dedicated tooling. One obvious example would be a tool that could support the \DarTwin evolution process of applying a generic evolution transition pattern.

\subsection{The challenges of tooling}\label{sec:tooling}
Tooling was one of the drivers for applying \SysML. Since \DarTwin was originally a graphical notation, it was important to determine whether the tools could render the \DarTwin textual models into something resembling the \DarTwin notation. The available tools that we tested had different issues.

Let us preface the remaining discussion by stating that tool support for \SysML is still in its early stages. We do not intend to criticize these tools; rather, we merely describe our findings in testing them. We are in touch with the developers to see if some of our issues can be resolved.

We have looked at three \SysML tools:

\begin{itemize} 
   \item \SysML pilot implementation - Eclipse Plugin \cite{SysMLv2Pilot}
    \item Tom Sawyer \SysML Viewer v1.1.1 \cite{TomSawyerBeta}
    \item SysON V6.0 \cite{SysONBeta}
     
\end{itemize}

Common to all automatic renderings that we have seen is that they do not comply with the principles of placements that the \DarTwin notation demands: goals at the top and the twin system composite structure at the bottom and we did not find mechanisms that could provide that. 

\paragraph{\SysML Pilot Implementation}\Cref{fig:strawberry-dartwin-pilot} shows a rendering of the \SysML pilot implementation. The pilot implementation bases its graphic renderings on PlantUML\footnote{\url{https://plantuml.com/}} which is unaware of the specific \DarTwin notation. PlantUML is also a textual notation, but the version used in the Pilot Implementation is such that the graphics cannot be changed manually. Furthermore, the outdated version of PlantUML used to generate the diagrams contains some placement bugs that are detrimental to showing \DarTwin. 

Additionally, we ran into an issue on how the pilot implementation handled some specific keywords. We apply the language extension mechanisms to create our own keywords for \DarTwin. This is useful because it makes it very clear what is \DarTwin and what is the general \SysML. The keywords are defined through metadata as shown in \Cref{fig:xxx0} and the keywords are related to concepts defined in basic \SysML. We defined the \DarTwin concept \textsf{Goal} to be a \SysML \textsf{Requirement}. Within a \textsf{Requirement}, it is commonplace to define "require constraint", but we get a syntax error on that in the pilot implementation. The reason is that for simplicity of the compiler, some constructs are related syntactically. While our goals are semantically requirements, they are not syntactically requirements since we are using our own keyword \dartwinInline{\#goal}.

\paragraph{Tom Sawyer} \Cref{fig:strawberry-dartwin-TS} shows a rendering of the Tom Sawyer \SysML Viewer, where we manually arranged the graphic elements similar to the \DarTwin notation. 
Despite this reorganization, it still looks quite different and does not provide the same intuition as the \DarTwin notation. 
Furthermore, the tool struggles with the \SysML language extension features, especially the keywords used to define \DarTwinDSL and the representation of goals. 
Specifically, goals were sometimes imported into incorrect elements rather than their designated core, indicating a random assignment during the import process.

\paragraph{SysON} SysON \Cref{fig:syson}, an editor that allows manual generation and placement of elements,  proved difficult due to its poor parsing of the \DarTwinDSL syntax and inability to represent key elements such as goals in interconnection diagrams. 
The version we used, while functional, often produced inconsistent diagram layouts, and also struggled with the \SysML language extension keywords used to define \DarTwin. In the end, we generated most of the graphic elements from scratch. However, instead of proper connections, the exported model only declared actions without connecting any elements. As a result, the export to text failed to produce a meaningful \SysML description.

\paragraph{The Current State}
Despite these observations, there are many indicators pointing at tools' near-term improvements. 
There are several tool vendors working on \SysML tools. 
To mitigate the current lack of a standard for storing or exchanging graphic notation data, there is a group within the System Modeling Community (organized by the OMG) dedicated to defining this exchange format. 
\SysML further foresees mechanisms for the end user-defined graphical rendering. As of now, however, there are only a small number of diagram views declared, and thus implemented by the vendors.

\section{Related Work}
\label{sec:related-work}

\subsection{Describing System and Software Evolution}
In this paper we have concerned ourselves with \DarTwin - a language to describe evolutions of \DTSs. We have been concerned with precision as well as tooling and have explored how \SysML could help.
Evolution is a process, and as such we could have applied any behavioral language, but our aim is limited to \DTSs. Still there are related fields of change that have resemblance to our approach.
The discipline of self-adaptive systems~\cite{de2013software}, for instance, aims to automate the reactive system re-configuration to address various changes, with MAPE-K loops~\cite{kephart2003vision} being among the best-studied techniques. 
Software system evolution has led to Lehman's well-known \emph{eight laws}~\cite{lehman1996laws}, and has since been shown to be also applicable in software-intense and \aclp{cps}~\cite{VOGELHEUSER201554}.
Clearly, the \devops of \DTs can also be seen as natural evolution and has been studied in various ways in~\cite{combemale2023model,heithoff2024model}. 
Some research are more specific about \DT evolution.  \cite{David.Bork_2023} and \cite{Michael.etal_2024} proposed a general taxonomy framework. \cite{Aissat.etal_2024} implement an \DT architecture framework that aims to support \DT evolution with the help of DevOps. The former two do not seek operationalization and the latter one does not explore any evolution scenarios per se.

\subsection{MDE for Evolution, Evolution for Models}
Variability and Evolution has also been addressed in \ac{mde}-based domains in various forms. 
\Acp{dspl}~\cite{4488260} have been proposed to enables the binding/reconfiguration of variation points of SPLs at runtime. 

\SysML natively supports the expression of \emph{snapshots}~\cite{jansen2022language}, which may be used to describe a system's state at a given time, comparable to the \dartwinInline{\#dartwin\_before} and \dartwinInline{\#dartwin\_after} (see \eg~\cite{gomez2018temporalemf} and~\cite{bill2018need}). The concept, however, is limited to small-scale variations of property values, rather than large system reconfigurations.

Closely related to our work's changes in purposes, we can look towards \acp{adl}. \cite{goknil2016rule}, for instance, studied automated evolution of AADL models based on requirements changes, and \cite{ten2009change} studied the impact of changes in \SysMLone requirements.

Model-driven engineering often relies on model transformations. It allows for transforming one model into another for various purposes~\cite{Lucio2014}. In our contribution, we employ a manual approach to model transformation, applying the patterns to our model. Automated techniques are plentiful in the literature. For example, \cite{Czarnecki2006} provides an overview of the features of model transformation languages. For SysML v2, no dedicated graph-based transformation language is available; therefore, we applied the patterns manually. 

For creating such a language, we can look to the work introducing T-Core~\cite{Syriani2013}. T-Core shows a collection of transformation primitives where one is the pre- and post-condition patterns allowing to specify an evolution pattern, as in our \DarTwinDSL. In our case, the engineer selects the correct DarTwin as the starting point manually. The matching and rewriting of the pattern is done by applying the manual 5-step procedure. However, if automation is required, we need to consider more of the transformation primitives.  

As SysML v2 is text-based, it also opens up avenues for using text-based transformation languages, such as a template-based approach

\subsection{\acp{dsl}, Profiles, Formalization}
From a different viewpoint, \DarTwinDSL relates to the development of \DSLs, using \SysML as host language, using the natively provided means and mechanisms for customization.
The concept of embedding \DSLs in other host languages has been advocated for before~\cite{selic2007systematic}. UML, for instance, uses \emph{profiles}~\cite{fuentes2004introduction} for adjusting the syntax and semantics.  Unlike with \DarTwin in \SysML, the semantics of UML profiles were not defined formally through UML.
Notably, AADL further provides an \emph{annex} mechanism that allows itself to be extended to add, \eg discrete behaviour~\cite{FrancaAssessmentAADLBehavioral}, continuous behaviour~\cite{ahmad2014hybrid} or error modelling capabilities~\cite{delange2014architecture} to the language. 
This concept can also be extended to \emph{internal} and \emph{embedded} \DSLs, which ``use, and abuse''~\cite{freeman2006evolving} programming languages as hosts, as shown by SystemC~\cite{Black:2005:SGU:1197604} and CREST~\cite{Klikovits:2020:sosym:crest}, which use C++ and Python, respectively.

\SysML has already been targeted as host language for domain-specification. In~\cite{10350780}, the authors embed variability modelling capabilities within \SysML. Like their variability DSL, our evolution-focused \DarTwinDSL applies SysML v2 extensibility
to support precise modelling and systematic reuse. \cite{aadlv2lib} and \cite{ahlbrecht2024exploring} explore the creation of domain-specific libraries.
Despite its extensibility, \SysML presents limitations that are directly relevant to our work. \cite{jansen.etal2022} analyzes the language’s grammar and tooling, identifying gaps in modularity, semantics, and variant handling. These issues motivate our decision to define a custom DSL within \SysML rather than relying solely on profiles or annotations. 

\section{Conclusion \& Future Work}
\label{sec:conclusion}
\glsresetall

This paper describes the creation of a \DarTwinDSL based on \SysML. Our work has allowed the discovery of inaccuracies in the original \DarTwin publication \cite{Mertens2024} that resulted from the more informal starting point described.

While formalizing the \DarTwin notation using \SysML constructs for user-defined keywords was fairly simple, we further noticed \DarTwin's lack of clear concept for a transformation. Thus, we defined the description of \DTS evolutions as a set of three interrelated \DarTwin models: 
\dartwinInline{\#dartwin\_core} represents the stable part that is unaffected by the evolution, 
the elements that need removing are specified as \dartwinInline{\#dartwin\_before} (which specializes the core), and \dartwinInline{\#dartwin\_after} describes the elements that are added. It also specializes the core. We also provide a new notation form that merges these three \textsf{DarTwins} in a single \dartwinInline{\#dartrans} diagram.

Next, to facilitate the evolution workflow, we describe a \emph{5-step procedure} for the application of evolution patterns.

Through the two use cases used for validation, we learned that the transition patterns defined in ~\cite{Mertens2024} were not sufficient and general enough when applying \DarTwin in a more formal context.

The development of the \DarTwinDSL further enabled identification of numerous problems with the current state of \SysML's tooling, especially the graphical view definitions. Some of the problems were just general bugs that we would expect will be fixed within short time. Other issues were directly related to recreating the \DarTwin notation automatically. 

Our exploration discovered the following requirements for future development of the DarTwin DSL::
\begin{itemize}
    \item Increase, enhance and generalize the \DarTwin transition pattern library.
    \item Provide tooling to render the \DarTwinDSL into \DarTwin notation.
    \item Provide tooling to support using the \DarTwin procedure for evolution, most desirably with graphic interaction.
\end{itemize}

\bibliographystyle{IEEEtran}
\bibliography{bibliography}

\end{document}